\documentclass[a4paper,11pt]{article}
\pdfoutput=1 

\usepackage[T1]{fontenc} 
\usepackage{graphicx}
\usepackage{caption}
\usepackage{subcaption}
\usepackage{siunitx}
\usepackage{xcolor}
\usepackage{gensymb}
\usepackage{jcappub} 

\newcommand\lya{Ly$\alpha$}
\newcommand\hi{H\,{\sc{i}}}
\newcommand\mpc{\mathrm{h^{-1} Mpc}}
\newcommand\gpcvol{\mathrm{h^{-3} Gpc^{3}}}
\newcommand\mpcvol{\mathrm{h^{-3} Mpc^{3}}}

\title{\boldmath A tomographic map of the large-scale matter distribution using the eBOSS - Stripe 82 \lya~forest}

\author[a,1]{C. Ravoux\note{Corresponding author.}}
\author[a]{E. Armengaud}
\author[a]{M. Walther}
\author[a]{T. Etourneau}
\author[a]{D. Pomar\`ede}
\author[a]{N. Palanque-Delabrouille}
\author[a]{C. Y\`eche}
\author[b]{J. Bautista}
\author[c]{H.~du~Mas~des~Bourboux}
\author[a]{S. Chabanier}
\author[c]{K. Dawson}
\author[a]{J.-M. Le Goff}
\author[d]{B. Lyke}
\author[d]{A. D. Myers}
\author[e]{P. Petitjean}
\author[f]{M. M. Pieri}
\author[a]{J. Rich}
\author[g]{G. Rossi}
\author[h,i]{D. P. Schneider}

\affiliation[a]{IRFU, CEA, Universit{\'e} Paris-Saclay, D36, Gif-sur-Yvette F-91191, France}
\affiliation[b]{Institute of Cosmology \& Gravitation, University of Portsmouth, 1-8 Burnaby Road, Dennis Sciama Building, Portsmouth, PO1 3FX, U.K.}
\affiliation[c]{Department of Physics and Astronomy, University of Utah, 115 S 1400 E, Salt Lake City, UT 84112, U.S.A.}
\affiliation[d]{Department of Physics and Astronomy, University of Wyoming, 1000 E University Ave, Laramie, WY 82071, U.S.A.}
\affiliation[e]{Institut d'Astrophysique de Paris, Sorbonnes Universit\'es and CNRS, 98bis Boulevard Arago, Paris 75014, France}
\affiliation[f]{Laboratoire d'Astrophysique de Marseille, CNRS, CNES, Aix-Marseille Universit{\'e}, 38 Rue Fr{\'e}d{\'e}ric Joliot Curie, Marseille 13013, France}
\affiliation[g]{Department of Physics and Astronomy, Sejong University, Neungdong-ro, Gunja-dong, Gwangjin-gu, Seoul 143-747, Korea}
\affiliation[h]{Department of Astronomy and Astrophysics, The Pennsylvania State University, 525 Davey Lab, University Park, PA 16802, U.S.A.}
\affiliation[i]{Institute for Gravitation and the Cosmos, The Pennsylvania State University, 525 Davey Lab, University Park, PA 16802, U.S.A.}

\emailAdd{corentin.ravoux@cea.fr, eric.armengaud@cea.fr, michael.walther@cea.fr, thomas.etourneau@cea.fr, daniel.pomarede@cea.fr, nathalie.palanque-delabrouille@cea.fr, christophe.yeche@cea.fr, julian.bautista@port.ac.uk, h.du.mas.des.bourboux@utah.edu, solene.chabanier@cea.fr, kdawson@astro.utah.edu, jean-marc.le-goff@cea.fr, blyke@uwyo.edu, geordiemyers@gmail.com, ppetitje@iap.fr, matthew.pieri@lam.fr, james.rich@cea.fr, graziano@kias.re.kr, dps7@psu.edu}

\abstract{
The Lyman-$\alpha$ (hereafter \lya) forest is a probe of large-scale matter density fluctuations at high redshift, $z > 2.1$. It consists of \hi~absorption spectra along individual lines-of-sight. If the line-of-sight density is large enough, 3D maps of \hi~absorption can be inferred by tomographic reconstruction. In this article, we investigate the \lya~forest available in the Stripe 82 field ($220\,\mathrm{deg^{2}}$), based on the quasar spectra from SDSS Data Release DR16. The density of observed quasar spectra is $37\,\mathrm{deg^{-2}}$ with a mean pixel signal-to-noise ratio of two per angstrom. This study provides an intermediate case between the average SDSS density and that of the much denser but smaller CLAMATO survey. We derive a 3D map of large-scale matter fluctuations from these data, using a Wiener filter technique. The total volume of the map is $0.94\,\gpcvol$. Its resolution is $13\,\mpc$, which is related to the mean transverse distance between nearest lines-of-sight. From this map, we provide a catalog of voids and protocluster candidates in the cosmic web. The map-making and void catalog are compared to simulated eBOSS Stripe 82 observations. A stack over quasar positions provides a visualization of the \lya~forest-quasar cross-correlation. This tomographic reconstruction constitutes the largest-volume high-redshift 3D map of matter fluctuations.
}
\arxivnumber{2004.01448}

\begin{document}

\maketitle
\flushbottom

\section{Introduction}
\label{sec:Intro}

Measuring the distribution of matter in the Universe is a major goal of observational cosmology. Features in this distribution provide information about the energy and matter content of the Universe as well as its gravitational evolution. The \lya~forest is a key probe of this distribution. It consists of absorptions in the electromagnetic spectrum of bright and distant sources such as quasars, due to the \lya~transition of intervening neutral hydrogen located at various redshifts along their lines-of-sight. As such, it provides a one-dimensional measurement of the amount of neutral hydrogen in the Intergalactic Medium (IGM)~\cite{Croft1997}. In this medium, the balance between processes such as photo-ionization by the ultraviolet background and electron recombination links the abundance of neutral hydrogen to the total baryon density. The latter is in turn correlated on large scales with the total matter distribution.

Quasars have been detected out to redshifts of 7.5 \cite{2018Bados}. Ground-based observations provide \lya~forest measurements above $z=2.1$. The \lya~forest is one of the main probes of large-scale matter distribution at high redshift. Constraints on cosmological models, as well as on the neutrino mass, were obtained from the relatively small-scale 1D measurements of \lya~forest absorption~\cite{Seljak2006,Palanque2015a,PalanqueDelabrouille2015,Yeche2017,PalanqueDelabrouille2019}. Recently, the increase of both the line-of-sight density and observational footprint in large sky surveys allowed the measurement of 3-dimensional \lya~forest correlations~\cite{Slosar2011}. In particular, the \lya~forest was used to measure the Baryon Acoustic Oscillation (BAO) scale through auto-correlation and cross-correlation with quasars \cite{Busca2013,Slosar2013,Font-Ribera2013,Delubac2014,Bautista2017,Bourboux2017,Blomqvist2019}.

While the \lya~forest data is one-dimensional by nature, it is possible to construct 3-dimensional maps from 1D data on scales comparable to the mean separation between lines-of-sight \cite{Petitjean1997}. Such an endeavour was already performed by the COSMOS \lya~forest Mapping And Tomography Observations (CLAMATO) survey~\cite{Lee2013,Lee2014,Lee2018}. This program uses a dense set of spectroscopic observations from Lyman-break galaxies (LBGs) to reconstruct the cosmic web in a small area ($0.157\,\mathrm{deg^{2}}$). With a redshift range $z= 2.05 - 2.55$, this survey provides a comoving volume of $30 \times  24 \times   438\, \mpcvol$. With the combined use of LBGs and quasars, the achieved line-of-sight density is $1455\,\mathrm{deg^{-2}}$. The ultimate purpose of such high sampling programs is to outline the cosmic structure down to the scale of filaments, i.e.~of the order of a Mpc. Accessing filament scales would be possible with future observations using e.g.~the Extremely Large Telescope (ELT)~\cite{Japelj2019}. The performance of the algorithm used by CLAMATO was assessed through cosmological simulations in~\cite{Cisewski2014,Ozbek2016}. These studies determined in particular the map-making performance for different mean separation $\langle d_{\bot} \rangle$ between individual lines-of-sight.

In this article, we present a high-redshift \lya~forest absorption map on much larger scales, from part of the quasar spectra obtained by the Sloan Digital Sky Survey-IV (SDSS-IV) \cite{SDSS42016,SDSS42017,2013AJ....146...32S,SDSS2006}. They are provided by the 16$^{th}$ Data Release (DR16)~\cite{DR162019} of the SDSS Extended Baryon Oscillation Spectroscopic Survey (eBOSS). We use the same tomographic reconstruction algorithm developed by CLAMATO. The authors in \cite{Cisewski2014} also exploited BOSS DR9 data to obtain a tomographic map, but with a lower line-of-sight density than in this study. Here we focus on the Stripe 82 part of the SDSS footprint (right ascension from 317\degree to 45\degree and declination from -1.25\degree to +1.25\degree).

\lya~forest tomographic mapping is already used for several applications. It has recently been applied to detect an enormous \lya~nebula on a very small portion of the sky~\cite{Mukae2019} centered on the \lya~forest protoclusters found by the MAMMOTH project~\cite{Cai2016,Cai2017}. \lya~forest tomography was also used to identify underdensities and overdensities in the matter distribution of the Universe. As an example, a $z=2.45$ protocluster was discovered in \cite{Lee2015} and later confirmed as an overdensity of spectroscopically-observed galaxies. We perform a similar search with the tomographic map obtained in this article, and provide the most probable candidates for very large protocluster systems in the Stripe 82 field.

By applying a similar method to underdensities, large voids can be detected in the cosmic web. The statistical properties of voids provide an interesting probe for cosmology. For example, galaxy-based void catalogs were used, together with galaxies themselves, to improve Redshift Space Distortion (RSD) measurements with SDSS data \cite{Nadathur2015a,Nadathur2015b,Nadathur2019}. One can produce an Alcock \& Paczynski cosmological consistency test \cite{Alcock1979} by stacking voids and studying their profile (see e.g.~\cite{Lavaux2012,Demchenko2016}).
While several methods were developed to detect voids from galaxy surveys, detection based on the \lya~forest is less explored. It is possible to search for voids from the one-dimensional \lya~forest data, as shown in~\cite{Viel2008}, but void searches based on 3D maps are clearly more efficient and robust, making use of redundant information from nearby lines-of-sight. A void search method based on tomographic \lya~forest mapping was developed in~\cite{Stark2015a} and applied in~\cite{Krolewski2018} to produce a void catalog from the cosmic web. In this article, we apply a similar method to produce a catalog of void candidates from the Stripe 82 tomographic map. Given the volume and resolution of this map, we can retrieve large voids in the cosmic web.

This article is laid out as follows: Sect.~\ref{sec:Stripe} describes \lya~forest spectra available in eBOSS DR16, in the Stripe 82 footprint. Sect.~\ref{sec:Algo} presents the method used to create our tomographic map. The quality of the map-making is assessed with dedicated simulations in Sect.~\ref{sec:Mocks}. We describe methods to search for under- and overdensities in Sect.~\ref{sec:Underdensities}. Sect.~\ref{sec:Results} presents the tomographic map based on eBOSS data, together with its properties including the results of void and large overdensity searches, as well as the inferred stacked \lya~forest flux signal around quasars.

\section{Lyman-alpha forest data}
\label{sec:Stripe}

We use the \lya~forest region from quasar spectra available in the 16$^{th}$ Data Release~\cite{DR162019} (DR16) of the SDSS-IV eBOSS survey~\cite{SDSS42016,SDSS42017,2013AJ....146...32S,SDSS2006}. The sky density of these objects, which is much smaller than that of LBGs, such as used by CLAMATO, is the main factor limiting the resolution of our tomographic map. For this reason, while the full eBOSS footprint could be used in principle, we choose here to focus on the Stripe 82 field, a narrow band in the Equatorial plane lying within the Southern Galactic Cap. In a large fraction of this field, a particularly dense and uniform sample of quasar spectra were acquired. Indeed, repeated photometric observations over a period of about ten years leads to targeting quasars and rejecting stellar contaminants in a particularly efficient way. This efficient selection was accomplished by combining the time-variability information from observed lightcurves together with color measurements~\cite{PalanqueDelabrouille:2010wi,2015ApJS..221...27M}. This improved targeting efficiency brings the mean quasar density in the Stripe 82 field to $37\,\mathrm{deg^{-2}}$. In comparison, with the same redshift range, the mean quasar density of the whole eBOSS footprint in DR16 is approximately $20\,\mathrm{deg^{-2}}$.

\begin{figure}[!t]
    \centering
    \includegraphics[trim=0cm 0cm 0cm 0cm, clip=true,width = 150mm]{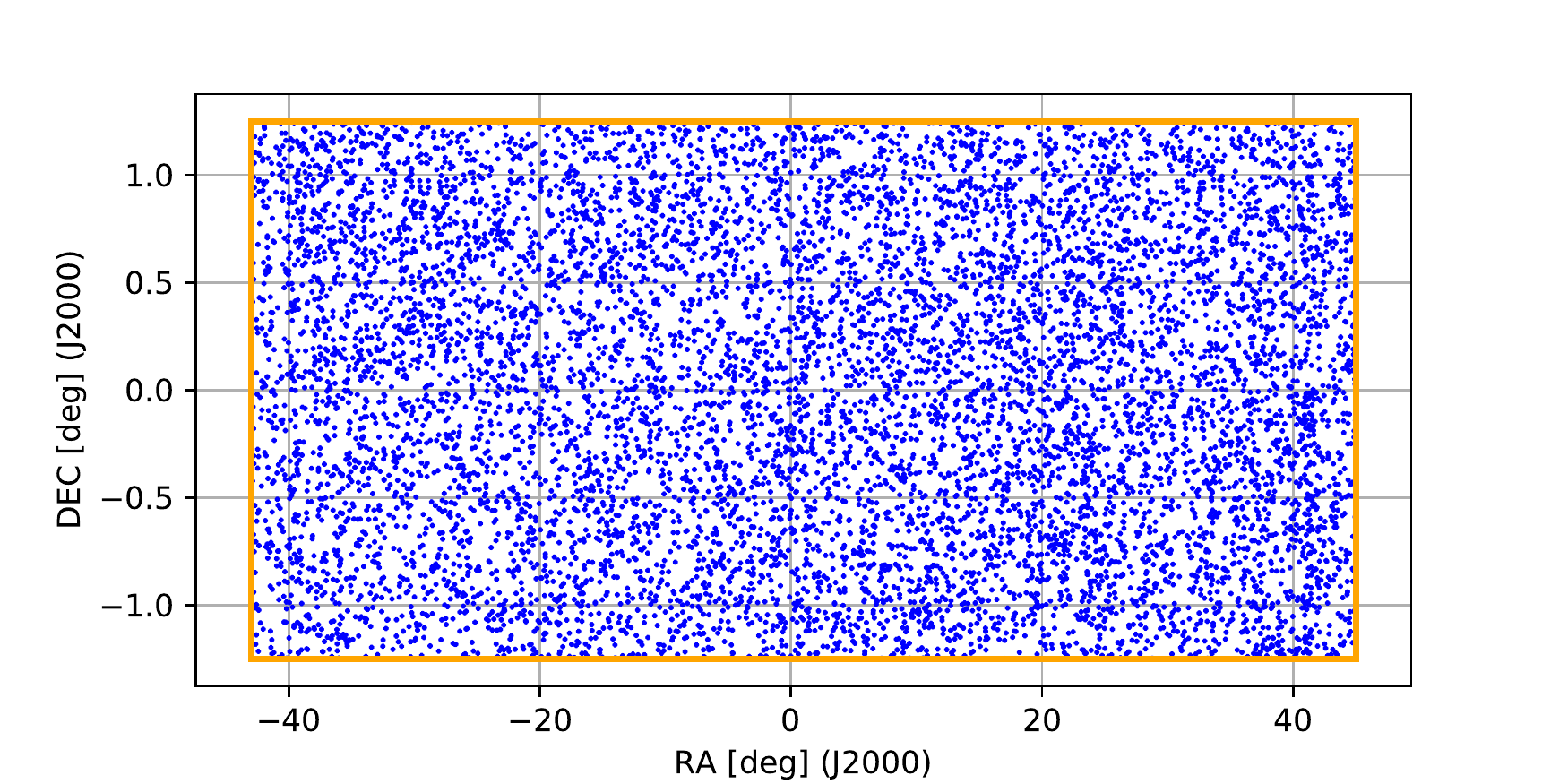}
    \caption{Sky distribution of the 8199 eBOSS - DR16 quasars in the Stripe 82 ($z = 2.1 - 3.2$), which provide \lya~forest spectra for this study. The total density is $37\,\mathrm{deg^{-2}}$. The scales of the horizontal and vertical axes are different.}
    \label{fig:ra_dec_map}
\end{figure}

\begin{figure}[!t]
    \centering
    \includegraphics[trim=0cm 0cm 0cm 0cm, clip=true,width = 130mm]{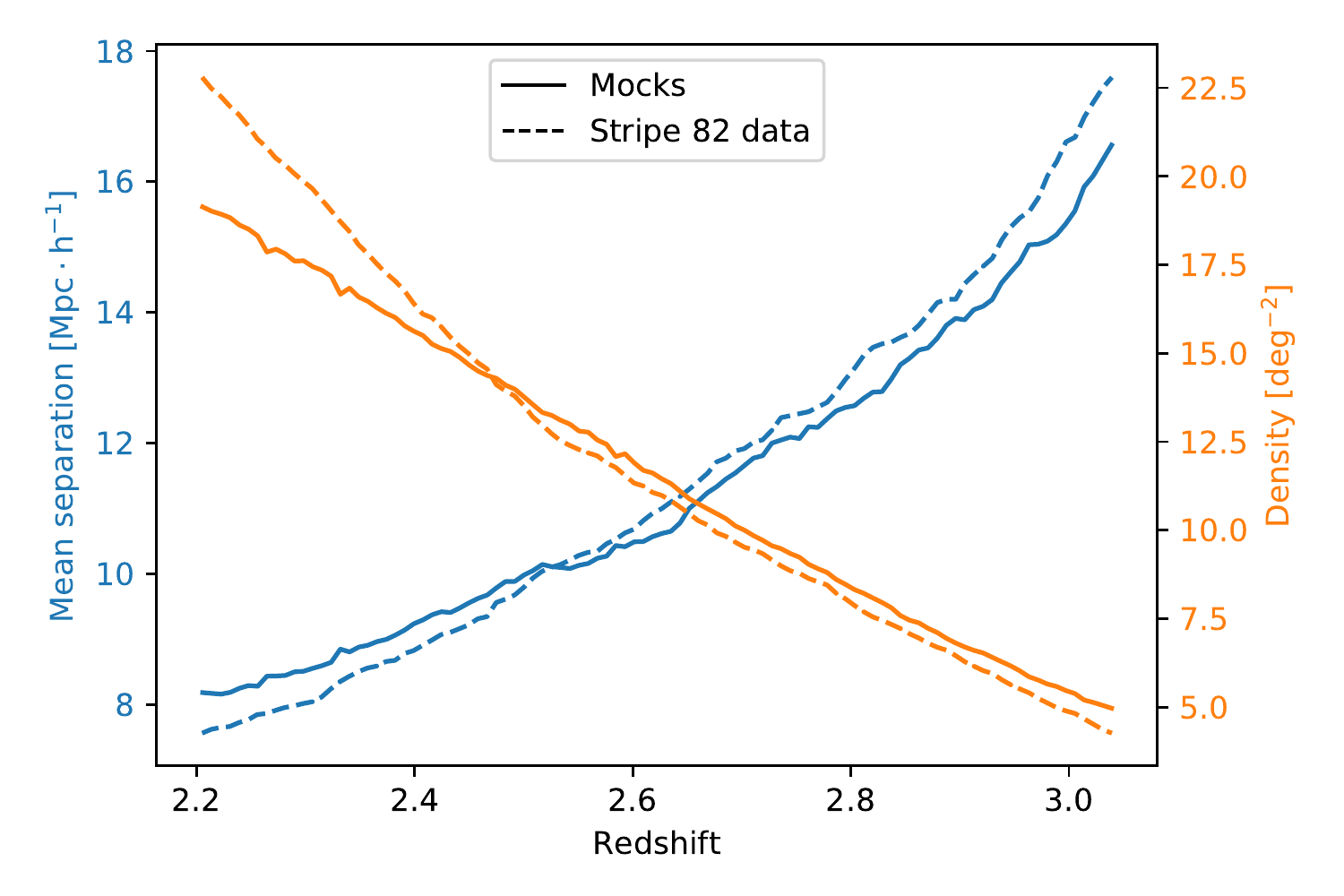}
    \caption{Blue: mean separation between nearest lines-of-sight [$\mpc$] as a function of redshift. Orange: line-of-sight effective density [$\mathrm{deg^{-2}}$], as a function of redshift. DR16 data are shown in continuous curves while dashed curves correspond to mocks presented in Sect.~\ref{sec:Mocks}. Higher difference at low redshift comes from the presence of high-noise quasars in the data which do not contribute quantitatively to the tomographic reconstruction. They are not modeled in the mocks.}
    \label{fig:separation}
\end{figure}

The footprint of our map within Stripe 82 is illustrated in Fig.~\ref{fig:ra_dec_map}. The right ascension range with a homogeneous \lya~forest dataset extends from -43\degree to +45\degree, and the declination covers Stripe 82 width, from -1.25\degree to +1.25\degree. This solid angle corresponds to $220\,\mathrm{deg^{2}}$. A total of 8199 \lya~forest spectra were obtained from quasar observations in this field, corresponding to the mean density of $37\,\mathrm{deg^{-2}}$. Lines-of-sight are selected over the redshift range $z = 2.1 - 3.2$. Fig.~\ref{fig:separation} shows the mean line-of-sight effective density as a function of redshift. It also represents the mean separation $\langle d_{\bot} \rangle$ between nearest lines-of-sight as a function of redshift.

Our quasar spectra originate from the SDSS DR16 Quasar catalog~\cite{dr16q}. Quasar targets were observed over the five-year period from 2009 to 2014 during the BOSS survey \cite{SDSS32011,SDSS32013} or between 2014 and 2019 during the SDSS-IV eBOSS survey \cite{SDSS42016,SDSS42017}. Spectra were obtained by co-adding individual observations from typically four 15-minute long exposures. The spectra pixel size is of $1\,\mathrm{\si{\angstrom}}$. An unbiased estimator for the observed flux $f(\lambda)$ is computed, weighting pixel data by the Charge Coupled Device (CCD) readout noise as in~\cite{Bautista2017,Bourboux2017}. For each spectrum, a \lya~forest flux contrast field $\delta_{F}(\lambda)$ (named flux contrast later) is defined by:

\begin{equation}
    \label{delta_flux}
    \delta_{F}(\lambda) = \frac{f(\lambda)}{C_{q}(\lambda)\overline{F}(z_{Ly\alpha})} - 1 . \\
\end{equation}
In this expression $C_{q}(\lambda)$ is the unabsorbed quasar flux (the continuum), while $\overline{F}(z_{Ly\alpha})$ is the mean fraction of flux transmitted by the IGM at the redshift $z_{Ly\alpha} = \lambda / \lambda_{\alpha} -1 $, such that by definition the average value of $\delta_{F}$ is zero at any redshift. The wavelength $\lambda_{\alpha} = 1215.668\,\mathrm{\si{\angstrom}}$ corresponds to the \lya~forest absorption peak at rest.

Following \cite{Bautista2017,Bourboux2017,Chabanier2018,Chabanier2019}, the \lya~forest flux contrasts are computed using the \texttt{picca}\footnote{\label{picca}Package for IGM Cosmological-Correlations Analyses, \url{https://github.com/igmhub/picca}} software, developed for the BOSS/eBOSS surveys. The product $C_{q}(\lambda)\overline{F}(z_{Ly\alpha})$ is evaluated directly with the same approach as in \cite{Bautista2017,Bourboux2017}. Although the Stripe 82 field is smaller than the full BOSS footprint, the number of quasars in our Stripe 82 sample is large enough to use this approach. The quasar continuum  is modeled as the product:

\begin{equation}
    \label{continuum}
    C_{q}(\lambda) = (a_{q} + b_{q}\log(\lambda))\,C\left(\lambda_{RF}=\frac{\lambda}{(1+z_{\rm quasar})}\right) ,
\end{equation}
where $a_{q}$ and $b_{q}$ are quasar-dependent normalization terms. The parameters $a_{q}$, $b_{q}$, and $C(\lambda_{RF})$ are determined iteratively by normalizing $C(\lambda_{RF})$ for each redshift and maximizing a likelihood function.

We restrict the \lya~forest flux contrast calculation to pixels for which the observed wavelength is in the range $3769 - 5106 \,\mathrm{\si{\angstrom}}$, corresponding to $2.1 < z < 3.2$. We also select the rest-frame wavelength in the range $1040 < \lambda_{RF} < 1200\,\mathrm{\si{\angstrom}}$, so that the measured flux fluctuations are dominated by the \lya~forest. In particular, we do not attempt to exploit the Lyman-$\beta$ forest at rest frame wavelengths below $1025\,\mathrm{\si{\angstrom}}$. The cut $\lambda_{RF} < 1200\,\mathrm{\si{\angstrom}}$ implies that for a given line-of-sight the closest selected pixel is located on average $49\,\mpc$ from the associated quasar. This decision mitigates most of the proximity effect: close to a quasar, the proportion of neutral hydrogen is indeed influenced by the large UV radiation background~\cite{Bajtlik:1988iy}.


In addition to the flux contrast $\delta_{F}$, \texttt{picca} provides an estimate of the instrumental noise $\sigma_{\delta_{F}}$. The corresponding signal-to-noise ratio per pixel $\mathrm{SNR} = F/\sigma_{F} = (\delta_{F} + 1)/\sigma_{\delta_{F}}$ is typically close to 2. This value is relatively large compared to the case of CLAMATO. One of the main reason of this difference is that SDSS quasars are considerably brighter than typical LBGs.

Broad Absorption Line (BAL) quasar spectra are not used. We use the balnicity index of the CIV absorption~\cite{Weymann1997,Trump2006}, removing quasars whose balnicity index is larger than zero. This suppress 6\% of quasars and is not counted in the 8199 quasars.

The flux contrast $\delta_F$ is also contaminated by High Column Density (HCD) objects such as Damped \lya~forest systems (DLAs)~\cite{Wolfe1986}. These typically originate from the lines-of-sight passing near foreground galaxies. Although HCD objects by themselves constitute tracers of the matter distribution, they have a specific bias and an extended impact on the observed spectra. For simplicity we therefore choose to mask the affected regions of the spectra.

To remove HCD contaminations, we use the DR16 HCD catalog obtained with the finder \texttt{dla\_cnn}~\cite{Parks2018}. This machine learning algorithm identifies candidate HCDs together with inferred \hi~column density and a detection confidence parameter. Among the candidates, we select those whose column density is $N_{\rm HI} > 10^{19.7} \, \mathrm{cm^{-2}}$ with a detection confidence larger than 70~\%. This limit corresponds to most DLAs, defined by  $N_{\rm HI} > 2 \times 10^{20}\, \mathrm{cm^{-2}}$~\cite{Wolfe1986}, and some Super Lyman Limit Systems, defined by $19 < \mathrm{log}_{10}(N_{\rm HI}/\mathrm{cm^{-2}}) < 20.3$~\cite{Prochaska2006}. In practice for our data set the HCD detection efficiency is poor for $\mathrm{log}_{10}(N_{\rm HI}/\mathrm{cm^{-2}}) < 20.3$. We remove the pixels where the estimated HCD-induced absorption is higher than 20~\%. The fraction of \lya~forest pixels lost by this cut is 11~\%. In addition, the absorption in the corresponding Lorentzian-profile damping wings that remain after the cut are corrected with a Voigt profile following \cite{Bautista2017}. The resulting set of ($\delta_{F}$, $\sigma_{\delta_{F}}$), as a function of sky coordinates and redshift, constitutes the input data to the next step of tomographic map-making.

\section{Tomographic map-making algorithm}
\label{sec:Algo}
\label{subsec:Wiener}

Tomographic reconstruction algorithms can be used to infer the 3-dimensional IGM density from a set of 1-dimensional \lya~forest line-of-sight. In \cite{Pichon2001,Caucci2008}, several techniques are presented to adapt tomographic methods to the creation of a 3D \lya~forest flux map. In \cite{Pichon2001}, a Bayesian method is presented. An advanced use of Bayesian methods in this context is also given in~\cite{Porqueres2019} and~\cite{Horowitz2019}, which apply a dynamic forward modelling approach to model the observed 1D lines-of-sight and reconstruct the 3D density and velocity fields of the IGM.

Under certain hypotheses detailled in~\cite{Pichon2001}, the Bayesian approach is mathematically equivalent to applying a Wiener filter \cite{wiener1949,Press1992}. Wiener filtering provides an unbiased minimum-variance linear estimator of a field. A simple implementation of this method, adapted to \lya~forest datasets, was developed by CLAMATO~\cite{Lee2013,Lee2014,Lee2018,Stark2015a,Stark2015b}; and our work uses the same public code, \texttt{DACHSHUND}\footnote{\label{clamato_algo}\url{https://github.com/caseywstark/dachshund}}. An advantage of this approach is its technical simplicity: it does not rely on complex assumptions for the underlying model, but still includes the effects of instrumental properties of the input data.

Since the mean separation between nearest lines-of-sight in our dataset is of the order of $10\,\mpc$, we aim to reconstruct a map with a resolution close to this value. Therefore only large-scale \lya~forest flux (or matter) fluctuations are retained in the reconstructed map. Their distribution is nearly Gaussian, so the underlying linear model used for Wiener filtering is well adapted to our case. In fact, \cite{Lee2018} found that Wiener filtering, with adapted parameters, is appropriate even in the case of the CLAMATO survey whose line-of-sight density is much larger than that of eBOSS.

The Wiener filter takes as an input the 1D line-of-sight data $\mathbf{d}$, assumed to be the sum of pixel signals $\mathbf{s}_{p}$ and Gaussian noise $\mathbf{n}$:

\begin{equation}
    \label{eq:signal}
    \mathbf{d} = \mathbf{s}_{p} + \mathbf{n}\ .
\end{equation}
The Wiener filter provides an estimate, $\mathbf{\hat{s}}= \mathbf{L}\cdot \mathbf{d}$, for the \lya~forest flux contrast signal over the entire volume by minimizing the difference between the reconstructed signal $\mathbf{\hat{s}}$ and the signal of the input pixels in the map called $\mathbf{s}_{m}$:

\begin{equation}
    \epsilon = E[|\mathbf{s}_{m} - \mathbf{\hat{s}}|^2]\ .
\end{equation}
The minimization, detailed e.g., in \cite{Pichon2001}, results in the expression for the operator $\mathbf{L} = \mathbf{S_{mp}}(\mathbf{S_{pp}} + \mathbf{N})^{-1}$. Here, the pixel-pixel and map-pixel covariance matrices $\mathbf{S_{pp}}$  and $\mathbf{S_{mp}}$ are both modeled by Gaussian kernels:

\begin{equation} 
    \label{eq:gaussian_kernel}
    \mathbf{S}_{ij} = \sigma_{F}^{2} \exp \left(- \frac{(r_{i\parallel} - r_{j\parallel})^{2}}{2 L_{\parallel}^{2}}       \right) \exp \left(- \frac{(r_{i\bot} - r_{j\bot})^{2}}{2 L_{\bot}^{2}}       \right),
\end{equation}
where $L_{\bot}$ and $L_{\parallel}$ are the transverse and longitudinal correlation lengths of the signal, which is discussed below. The parameter $\sigma_{F}^{2}$ is the expected variance of flux fluctuations in the map; in practice, it controls the amplitude of fluctuations in the reconstructed map. For our data, the noise matrix $\mathbf{N}$ is well approximated by a diagonal matrix with $\mathbf{N}_{ii} = n_{i}$, where $n_{i}$ is the individual pixel noise of pixel $i$.

The \texttt{DACHSHUND} public code makes use of a conjugate gradient Wiener filter. We convert the line-of-sight angular and redshift coordinates into pixel coordinates $(x,y,z)$ that represent angular, right ascension for $x$ and declination for $y$, and longitudinal, $z$, comoving distances, in $\mpc$. We assume the concordance $\mathrm{\Lambda CDM}$ cosmology with fixed $\Omega_{\Lambda} = 0.6853$, $\Omega_{m} =  0.3147$, according to \cite{Planck2018}. Since all coordinates are expressed in $\mpc$, no assumption on the value of $H_0$ is needed.
As mentioned in \cite{Lee2018}, the choice of cosmology only affects the global longitudinal and transverse scales. 

It is technically convenient to define the $(x,y,z)$ approximated coordinate system such that all lines-of-sight are parallel to each other, and the output map is a rectangular box. For that reason, comoving angular distances are computed for all pixels using the same redshift $z = 2.65$ corresponding to the middle redshift. The spread in declination of the input data is 2.5\degree, so that distance errors between the lowest and highest redshifts $z=2.1$ to 3.2 are of the order of a few percent along the $y$-axis of the map. In the right ascension direction, 88\degree~wide, the difference is much larger as the parallel approximation does not hold at all. Therefore, without any supplementary coordinate transformations, this map is not adapted for studies on large-scale distances.

The \texttt{DACHSHUND} input is a list of pixels vectors $(x,y,z,\delta_{F},\sigma_{\delta_{F}})$. It is used to construct $(r_{\parallel},r_{\bot},\mathbf{d},\mathbf{N})$ following the definition of the Wiener filter. The output is a 3D map containing the reconstructed \lya~forest flux $\delta_{\mathrm{Fmap}}$ that corresponds to the signal $\hat{s}$. The map is pixellized at approximately one pixel per $\mpc$.

Wiener filter tomographic reconstruction is not a parameter-free method. The main parameters are the normalization $\sigma_{F}$, and the transverse $L_{\bot}$ and longitudinal $L_{\parallel}$ smoothing lengths. The transverse $L_{\bot}$ correlation length is constrained by the fact that no transverse \lya~forest flux fluctuations can be reconstructed with a size smaller than the mean separation $\langle d_{\bot} \rangle$ between nearest lines-of-sight. Fig.~\ref{fig:separation} shows the evolution of this quantity as a function of redshift. The transverse correlation length increases with redshift in a way related to the redshift distribution of quasar sources used in the sample. From this figure, we choose $L_{\bot} = 13\,\mpc$, so that  $L_{\bot} > \langle d_{\bot} \rangle(z)$ for $z<2.8$, which covers most of the map. At high redshift, $z>2.8$, $L_{\bot}$ is smaller than the mean line-of-sight separation, a fact which must be considered when  interpreting the map.

In principle, the longitudinal smoothing length $L_{\parallel}$ could be set at a value close to a Mpc, which corresponds to the spectrograph resolution. However, since this value is much smaller than $\langle d_{\bot} \rangle$ we choose to set $L_{\parallel}=L_{\bot}=13\,\mpc$. This choice reduces the amplitude of anisotropies in the Wiener-filtered map, and keeps large-scale information while removing small-scale longitudinal fluctuations of the \lya~forest flux.

Another parameter to choose is the normalization factor $\sigma_{F}$, which controls the contrast of the map. The maximal reconstructed density contrasts are controlled by the ratio $\sigma_{F}^{2}/n_{\rm min}^{2}$, where $n_{\rm min}$ is the minimal value of the diagonal elements of the noise matrix $\mathbf{N}$~\cite{Stark2015b}. Indeed, in order to avoid numerical divergences in the Wiener filter algorithm, these matrix elements must be bounded by a floor value. Given the statistical distribution of pixel noise in our dataset, we adopt $n_{\rm min} = 0.1$, so that for the 6.3~\% of the pixels with lowest noise, this minimum value is used rather than the true pixel noise for the corresponding elements $n_i$. In the CLAMATO analysis, $n_{\rm min}$ was set to 0.2 due to larger relative noise in LBG spectra~\cite{Lee2018}. As for CLAMATO, we choose $\sigma_{F}^{2}$ such that the maximal contrast of the Wiener-filtered map is 1, which implies $\sigma_{F}^{2} = 0.01$.

Finally, for computational reasons, the map-making procedure is parallelized by dividing the Stripe 82 field into chunks, and merging the corresponding sub-maps at the end of reconstruction. To avoid edge effects, we allow the chunks to overlap, with an overlap width of $L_{\bot}$. This technique allows one to process large data samples without adding numerical artifacts.

The result of the tomographic reconstruction over the whole Stripe 82 is a map with $(5856\times180\times834)$ pixels. Its total size in terms of $(x,y,z)$ coordinates, in the right ascension, declination, redshift directions, is $(6352\times180\times835)\,\mpcvol$. The corresponding comoving volume, computed without parallel line-of-sight approximation, is $0.94\,\gpcvol$.

\section{Mocks and algorithm validation}
\label{sec:Mocks}

\subsection{Map-making on adapted simulations}
\label{subsec:mocks_map}

To assess the performance of tomographic reconstruction, we do not use physical simulations of the \lya~forest, as they require hydrodynamical effects and a resolution on the order of 100 kpc. No such simulation exists covering a near-Gpc$^3$ volume. Instead, we use much simpler synthetic data, called mocks, based on Gaussian random fields. Such mocks~\cite{Goff:2011cx,saclay-mocks,Farr2019} have been developed to reproduce the observed statistical properties of the \lya~forest on relatively large scales, over BOSS or DESI (Dark Energy Spectroscopic Instrument \cite{Desi2016}) footprints. These properties include the 3D auto-correlation function or the cross-correlation with other matter tracers, such as quasars.

We briefly summarize here the creation of these mocks. Gaussian fields are generated in a Fourier space box. The modes of this box are multiplied by several sets of appropriate weights such that, after applying an inverse 3D Fourier transform, they produce several boxes: a box of matter density fluctuations, $\delta_m$, at $z=0$; six boxes of velocity gradient components at $z=0$; and a box of quasar density fluctuations at $z=2.3$. A quasar catalogue is obtained by placing a quasar in each cell of this last box with a probability proportional to the logarithm of the density in the cell. 
For our study, this step was adapted in order to match the sky density and redshift distribution of eBOSS-DR16 quasars in Stripe 82, as described in Sect.~\ref{sec:Stripe}.
The \lya~forest transmitted-flux-fraction, $F$, along the line-of-sight to each quasar is finally obtained following the Fluctuating Gunn-Peterson \cite{Gunn1965} approximation, i.e., $F = \exp(-\tau)$, where the optical depth $\tau$ reads :

\begin{equation}
    \label{eq:flux_mocks_2}
    \tau = a(z)\exp\left[-b\,G(z)\,(\delta_{m}+\delta_{s} + c(z)\,\eta_{\parallel})\right] .
\end{equation}
In this expression, the field $\delta_{s}$ is added to $\delta_m$ in order to take into account small-scale flux fluctuations that are missing due to the relatively large, 2$h^{-1}$ Mpc, size of the box pixels. It is tuned to match the observed 1D power spectrum of the \lya~forest. The velocity gradient along the line-of-sight, $\eta_{\parallel}$, is obtained from the velocity-gradient component boxes. The parameters $a(z)$ and $c(z)$ are adjusted to match the observed \lya~forest bias and $\beta$ parameters. 
The parameter $c(z)$ controls redshift space distortions, and its value is actually close to $\beta$. $a(z)$ controls the bias. 
We set $b = 2-0.7(\gamma-1) = 1.58$, for an equation-of-state parameter of the intergalactic medium $\gamma=1.6$ \cite{HuiGnedin97}. Finally, $G(z)$ is the linear growth factor at the redshift of the considered pixel, with the convention $G(z=0)=1$. 

\begin{figure}[!t]
    \centering
    \includegraphics[trim=0cm 0cm 0cm 0cm, clip=true,width = 130mm]{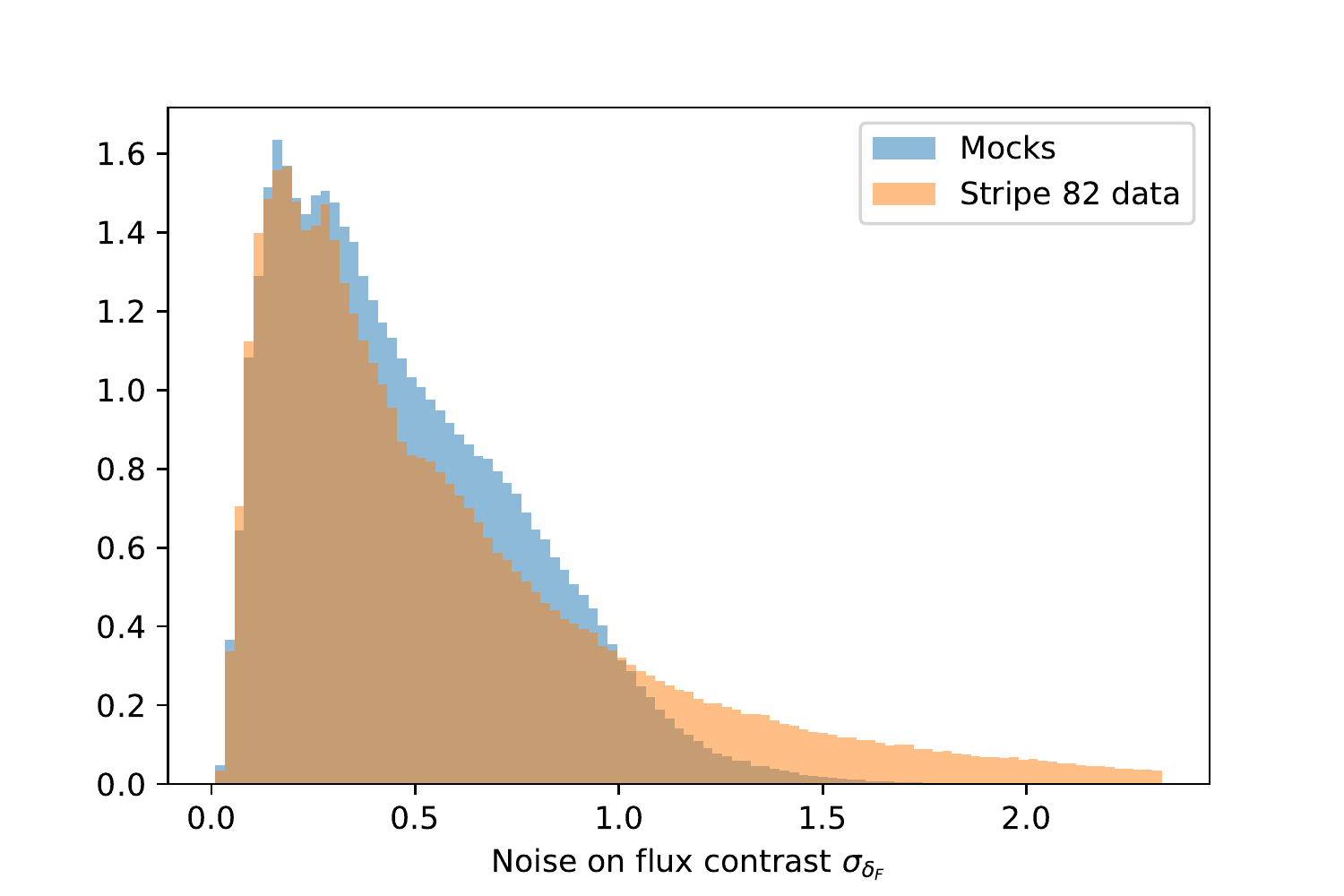}
    \caption{Normalized histograms of the flux contrast pixel noise $\sigma_{\delta_F}$ , for both our selected Stripe 82 data and mocks. A pixel corresponds to the flux contrast over $1\,\mathrm{\si{\angstrom}}$ width.}
    \label{fig:histogram_noise_comparison_mock_data}
\end{figure}

The flux is multiplied by a continuum to produce a quasar spectrum, and metals and HCD systems are added to the spectra using the \texttt{quickquasars}\footnote{\label{quickquasars}This code is developed for DESI simulations,  \url{https://github.com/desihub/desisim/blob/master/py/desisim/scripts/quickquasars.py}} program. HCD systems are then removed with the same method as for the data, as detailed in Sect.~\ref{sec:Stripe}.
The final outputs of the simulation are spectra and an associated quasar catalogue, over a footprint matching our Stripe 82 selection. BAL quasars are not included in mock spectra. The mock spectra are then treated with the same pipeline and parameters as the DR16 spectra detailed in Sect.~\ref{sec:Stripe}.

Figure~\ref{fig:separation} illustrates the agreement between mocks and data with respect to the line-of-sight separation as a function of redshift. The histogram of flux-contrast noise, $\sigma_{\delta_{F}}$, is presented for mocks and our selected data in Fig.~\ref{fig:histogram_noise_comparison_mock_data}. The average and standard deviations of the noise distributions are similar, but their shapes are significantly different, especially at low signal-to-noise ratio. This result is related to the fact that the number of high magnitude quasars is smaller for mocks than in the DR16 data. However, the difference in the high-noise tail has a minor impact on the tomographic map, because Wiener filtering assigns more weight to low-noise pixels. We checked this effect explicitly, by constructing maps for which high-noise data were removed. This result is in agreement with~\cite{Lee2018,Stark2015b}.

\begin{figure}[!t]
    \centering
    \includegraphics[trim=2.5cm 1cm 2cm 2cm, clip=true,width = 150mm]{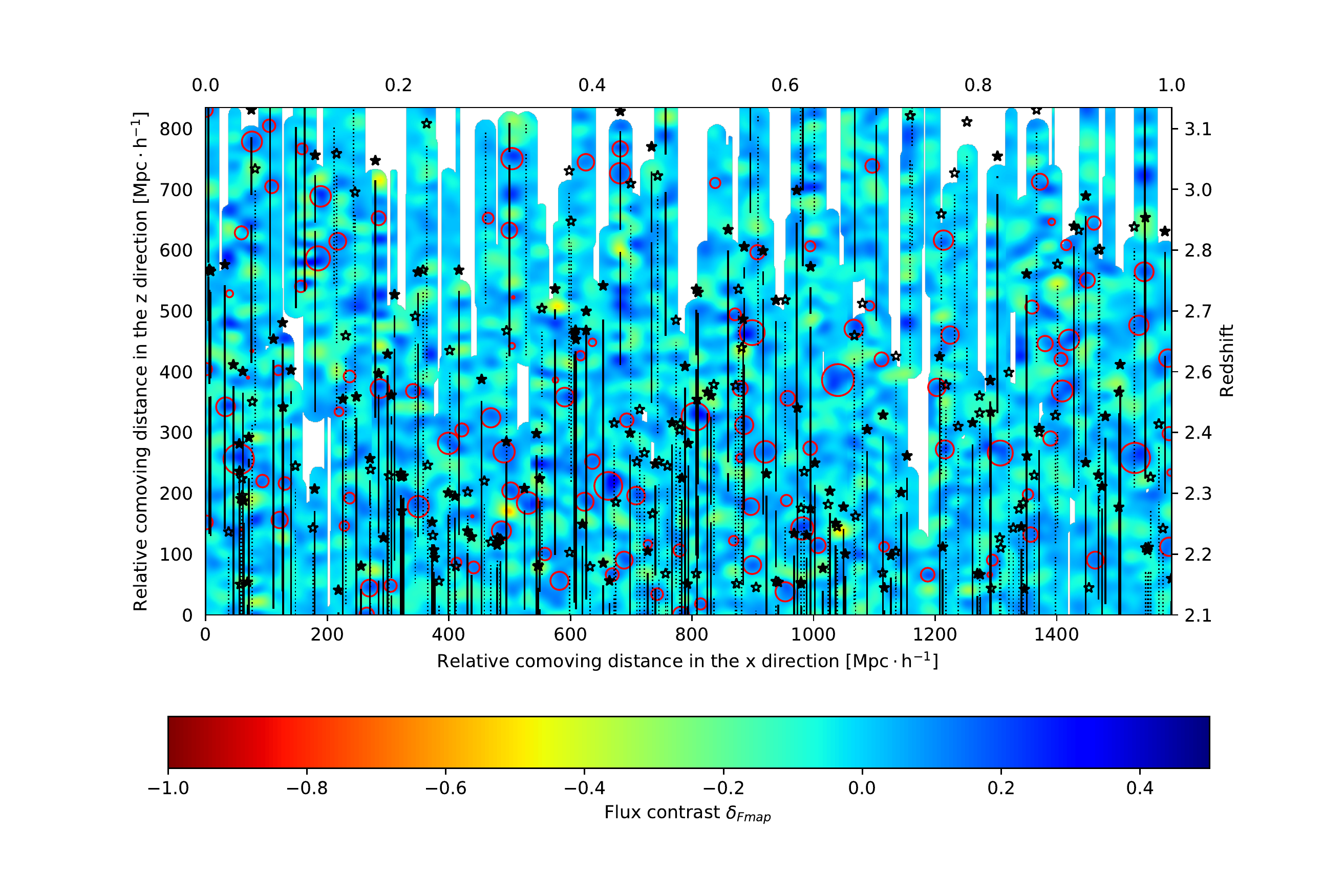}
    \caption{Slice through the tomographic \lya~forest flux map derived from mock data. It covers a 22\degree right-ascension range (along the $x$ axis), at fixed declination. The filtering parameters are described in Sect.~\ref{subsec:Wiener}. Circles are the intersection between the represented slice and identified voids. Filled (empty) stars represent quasars whose distance along the $y$ axis is less than 5 (10) $\mpc$ from the slice. Lines-of-sight used for the tomographic reconstruction are pictured as full (dotted) lines if they are 5 (10) $\mpc$ from the slice.}
    \label{fig:map_saclay_mocks}
\end{figure}

We apply the tomographic algorithm described in Sect.~\ref{subsec:Wiener} to all mock data. Fig.~\ref{fig:map_saclay_mocks} displays a slice of one quarter of the Stripe 82 mock map. 
The reconstructed \lya~forest flux contrast, $\delta_{\mathrm{Fmap}}$, is represented by a color map such that matter overdensities, corresponding to negative flux contrast, appear in red, while matter underdensities are in blue. Map pixels further than $20\,\mpc$ from any lines-of-sight are masked, as they can not be reasonably reconstructed; they appear in white in the representation.

\subsection{Performance of the matter density reconstruction}
\label{subsec:algo_perf}

Since the reconstructed map of \lya~forest flux contrast traces flux fluctuations on scales larger than $13\,\mpc$, it is expected to be an accurate representation of the associated matter fluctuations at similar scales. In order to estimate to what extent our tomographic map reproduces the underlying field of matter fluctuations, we use the reconstructed map of flux contrast $\delta_{\mathrm{Fmap}}$ from mock data and compare it to the input field of matter fluctuations $\delta_{\mathrm{m}}$. This random field was computed as a first step during mock data production. The field $\delta_{\mathrm{m}}$ is interpolated into the cartesian grid of the tomographic map. Since the original $\delta_{\mathrm{m}}$ was computed at $z=0$, it is multiplied by the linear growth factor at the considered redshift. Finally, $\delta_{m}$ is Gaussian smoothed with a smoothing length of $8\,\mpc$, which was chosen in order to optimize the correlation signal with the tomographic \lya~forest map.

\begin{figure}[!t]
    \centering
    \includegraphics[trim=0cm 0cm 0cm 0cm, clip=true,width = 130mm]{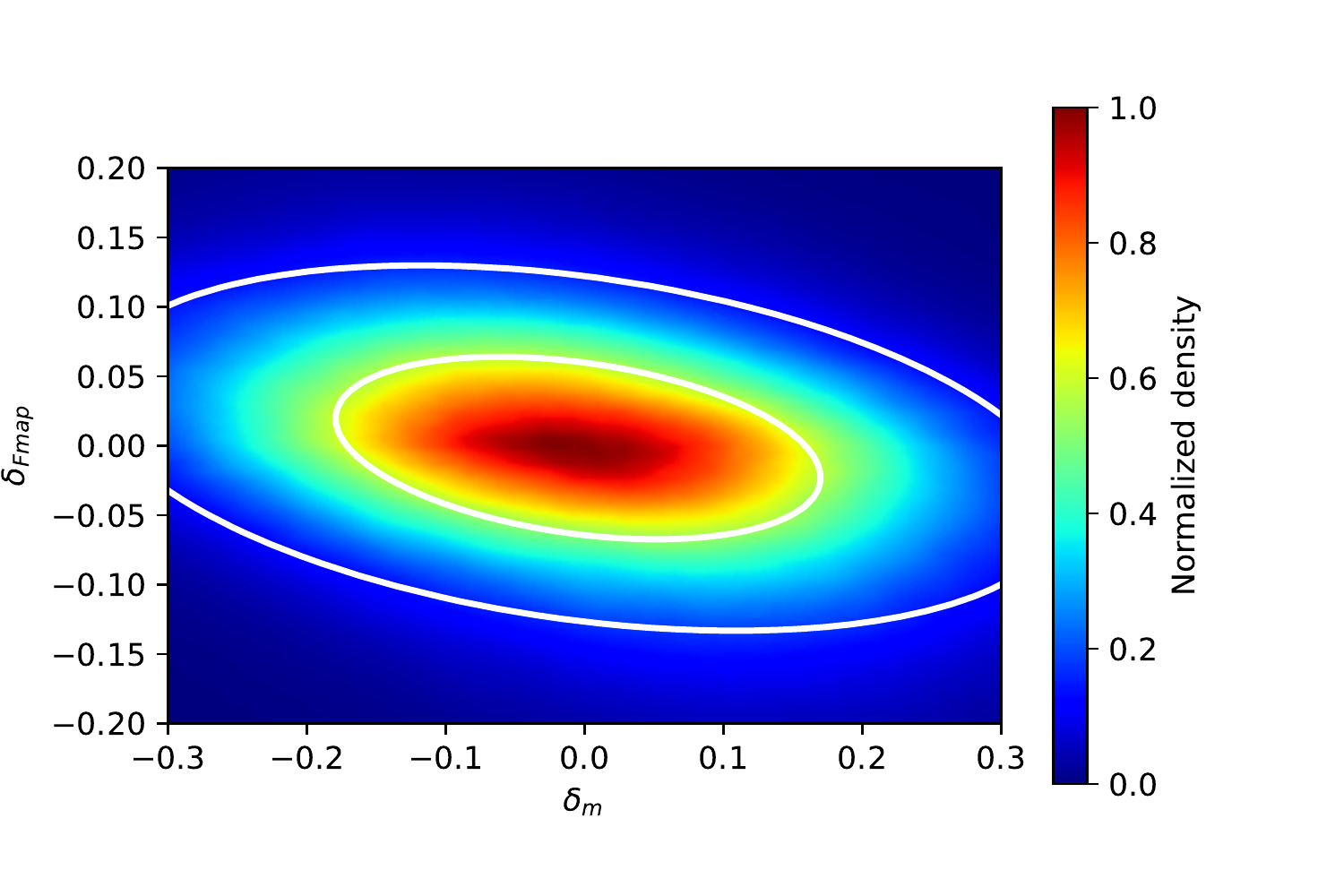}
    \caption{Comparison between the reconstructed flux contrast from our simulation and the input underlying field of matter fluctuations. The redshift range $z = 2.1 - 2.7$ is considered. The map of matter density fluctuations was Gaussian smoothed over $8\,\mpc$. White ellipses represent the $1\sigma$ and $2\sigma$ contours of a 2D Gaussian fit to the distribution.}
    \label{fig:comparison_density_flux}
\end{figure}

The resulting smoothed map of matter fluctuations is compared to the tomographic \lya~forest flux contrast map $\delta_{\mathrm{Fmap}}$ in Fig.~\ref{fig:comparison_density_flux}. At high redshifts, a large fraction of the map is masked, therefore we restrict the comparison to the redshift range $z = 2.1$ to $2.7$.

The correlation coefficient between both fields is $r = -0.34$. This number provides the simplest estimation of the ability of our tomographic approach to map the underlying cosmological field of matter fluctuations, and can be compared to simulation-based results of~\cite{Ozbek2016}. They tested several signal-to-noise ratios and line-of-sight separations and reported a correlation coefficient of $56\%$ for the closest values to our dataset ($\langle d_{\bot} \rangle = 12.65\,\mpc$ and signal-to-noise ratio of 2). Their correlation coefficient, however, was computed relative to the initial \lya~forest fluxes, not to the underlying density fields.
We can also compare our correlation factor with results from weak-lensing tomography of large-scale matter fluctuations. In particular, \cite{Oguri2017} used wide-field weak-lensing observations to construct a 3D tomographic map of matter fluctuations for $z = 0.1 - 1$ over a 167 deg$^2$ field. The comparison with matter density fluctuations as traced by galaxies yielded a correlation coefficient $r\sim 10$~\%.

As demonstrated in the simulations of~\cite{Ozbek2016}, the correlation of the tomographic map with the underlying matter field would naturally improve with smaller $\langle d_{\bot} \rangle$ and, to a smaller extent, larger signal-to-noise.



\section{Identification of underdensities and overdensities}
\label{sec:Underdensities}

Since the computed tomographic \lya~forest flux contrast map is anti-correlated with matter density fluctuations, an immediate application is to search for large under- and overdensities in the mapped volume.

\subsection{Overdensity search}
\label{subsec:protocluster_algo}

Protoclusters are defined as structures that will, at some stage, collapse into a galaxy cluster. At $z>2$, protoclusters are expected to be ubiquitous in the cosmic web, and to have a characteristic radius of 6~Mpc~\cite{Overzier:2016aat} for the largest ones. It is clear that, from the tomographic map produced in this study, only overdensities with a typical scale larger than $13\,\mpc$ can be identified. We may therefore expect, from our tomographic map, to be able to identify the largest protoclusters from $z = 2.1$ to $z = 3.2$ in the Stripe 82 field.
Based on the study of \cite{Stark2015b}, we designed a watershed search algorithm, similar to the one used by~\cite{Lee2015} to detect protoclusters in the CLAMATO survey.

We search for large deficit in the \lya~forest flux contrast to identify overdensities. In the watershed algorithm, all pixels whose $\delta_{\mathrm{Fmap}}$ is smaller than a given threshold, $\delta_{ws}$, are selected and then sorted into groups of neighbours. For each cluster, the center is defined as the pixel with the lowest $\delta_{\mathrm{Fmap}}$. We define the cluster radius by $R_{\mathrm{overdensity}} = (3N_{\mathrm{pixel}}V_{\mathrm{pixel}}/4\pi)^{1/3}$ where $N_{\mathrm{pixel}}$ is the number of pixels contributing to the cluster, and $V_{\mathrm{pixel}}$ is the pixel volume. In addition, clusters whose radius is smaller than a fixed parameter $R_{\mathrm{min}}$ are removed. The output of the algorithm is a catalog of overdensities characterized by a central position and a radius.

In the protocluster search presented in~\cite{Stark2015b}, the CLAMATO tomographic map is smoothed beforehand at a $4\,\mpc$ scale, matching the dimension of searched protoclusters. In our case, we are limited by the $13\,\mpc$ smoothing length of our map, so this step is not needed. 
The threshold $\delta_{ws}$ impacts both the positions and the radii of overdensities. As in~\cite{Stark2015a}, we set $\delta_{ws} = -3.5\,\Tilde{\sigma}_{\delta_{\mathrm{Fmap}}}$, where $\Tilde{\sigma}_{\delta_{\mathrm{Fmap}}}$ is the measured standard deviation of the tomographic map. The minimum cluster radius is set to $R_{\mathrm{min}} = 7\,\mpc$ so that identified overdensities have a diameter larger than the smoothing length of the map.  

Finally, in order to mitigate the impact of residual HCD systems and metals which could generate spurious overdensity detections (see e.g.~\cite{Cai2016}), we require that each selected cluster be crossed by lines-of-sight whose total length of crossing is larger than three times the cluster diameter. This criterion implies that each overdensity candidate is crossed by more than three lines-of-sight.

We stress again that only very extended overdensities in the cosmic web can be detected using our tomographic map. Interpreting their properties, in particular in terms of candidate protoclusters, would require dedicated N-body simulations, a task beyond the scope of this article. Applying our overdensity search procedure to the mock tomographic map results in the identification of nine candidates. Their mean radius is $10.5\,\mpc$. The values of the underlying matter density fluctuations, averaged within the overdensity radii, are positive for all candidates, with a mean value $\langle \delta_M \rangle = 0.4$.

\subsection{Void finding method}
\label{subsec:void_algo}

Voids are underdensities in the cosmic matter field located between filaments and walls. Their typical size is approximately $10\,\mpc$~\cite{Lavaux2012}, but they have a wide distribution of dimensions. With a tomographic map smoothed at a scale of $13\,\mpc$, we can identify large voids. The near-Gpc volume covered by Stripe 82 makes such a search particularly useful. The \lya~forest tomographic map is an opportunity to study the feasibility of void cosmology at high redshift with the \lya~forest~\cite{Stark2015b,Stark2015a}.

We adopt a simple-spherical algorithm to identify voids. The approach is more robust than the watershed algorithm with respect to noise on large scales as the watershed algorithm tends to provide complex void shapes~\cite{Stark2015a}.

The simple-spherical method selects all the pixels that have a flux contrast $\delta_{\mathrm{Fmap}}$ larger than a threshold $\delta_{th}$. At each selected point, a sphere is grown in radius until the mean flux value $\langle \delta_{\mathrm{Fmap}} \rangle$ inside the sphere reaches an average limit $\delta_{av}$. After treating all the selected pixels, some voids overlap. For each group of overlapping voids, the selection is done iteratively. The void with the largest radius is kept and all the spheres directly overlapping this void are suppressed. Groups of overlapping spheres are repeatedly created and the aforementioned procedure is applied until all remaining voids are separated. Voids with a radius smaller than a certain value $R_{\mathrm{min}}$ are deleted.

As for the watershed method, this algorithm relies on a few parameters on which the output void catalogs strongly depend. The parameter $\delta_{th}$ mainly controls the depth of the voids while $\delta_{av}$ controls the profile of the void, i.e., its size. The parameters $\delta_{av}$ and $\delta_{th}$ must be chosen simultaneously because of their correlated impacts on the void selection. If the value of threshold $\delta_{th}$ is too low, the centers of the detected voids are offset relative to the maximal $\delta_{\mathrm{Fmap}}$ inside the void. If $\delta_{th}$ is too high, only the deepest voids are detected. For reasonable values of $\delta_{th}$, modifying $\delta_{av}$ globally changes the void radius but does not move the center positions. Finally, $R_{\mathrm{min}}$ is used to select voids with a radius larger than the smoothing length of the map.

We adjusted the void detection parameters by applying the void finder to our mock tomographic map, in a similar approach to~\cite{Stark2015a}. The minimal radius is fixed at $R_{min} = 7\,\mpc$, identical to the minimal overdensity radius. Scanning over different algorithm parameters, we choose $\delta_{th} = 0.14$ and $\delta_{av} = 0.12$. With this choice the positions of large voids, of $\sim 25\,\mpc$ radius, are stable when varying the parameters around the central values.


Two additional criteria are applied to selected voids in order to remove those located in areas where the tomographic reconstruction is not able to operate without extrapolating. Similarly to the analysis of overdensities, we require that each selected void is crossed by lines-of-sight with a total length larger than the void radius. This constraint implies in particular that selected voids are crossed by at least two lines-of-sight. In addition, we require that for each selected void, less than 5~\% of its volume is located in the masked region, i.e., more than $20\,\mpc$ distant from any line-of-sight.
 
After applying these criteria, the void filling factor, defined as the ratio of the total void volume over the total map volume, is 14~\%, similar to the value of 15~\% found by~\cite{Stark2015a}.

\subsection{Properties of identified voids in mock data}
\label{subsec:underdensities_props}

The results of our void finding procedure on the mock tomographic map is presented in Fig.~\ref{fig:map_saclay_mocks}, where identified voids are represented by circles. The distribution of voids is rather uniform over the map. The largest reconstructed void radius from mock data is $34$~$\mpc$ as can be seen in Fig.~\ref{fig:histogram_comparison_mock_data_lin} (right).

\begin{figure}[!t]
    \centering
    \includegraphics[trim=0cm 0cm 0cm 0cm, clip=true,width = 130mm]{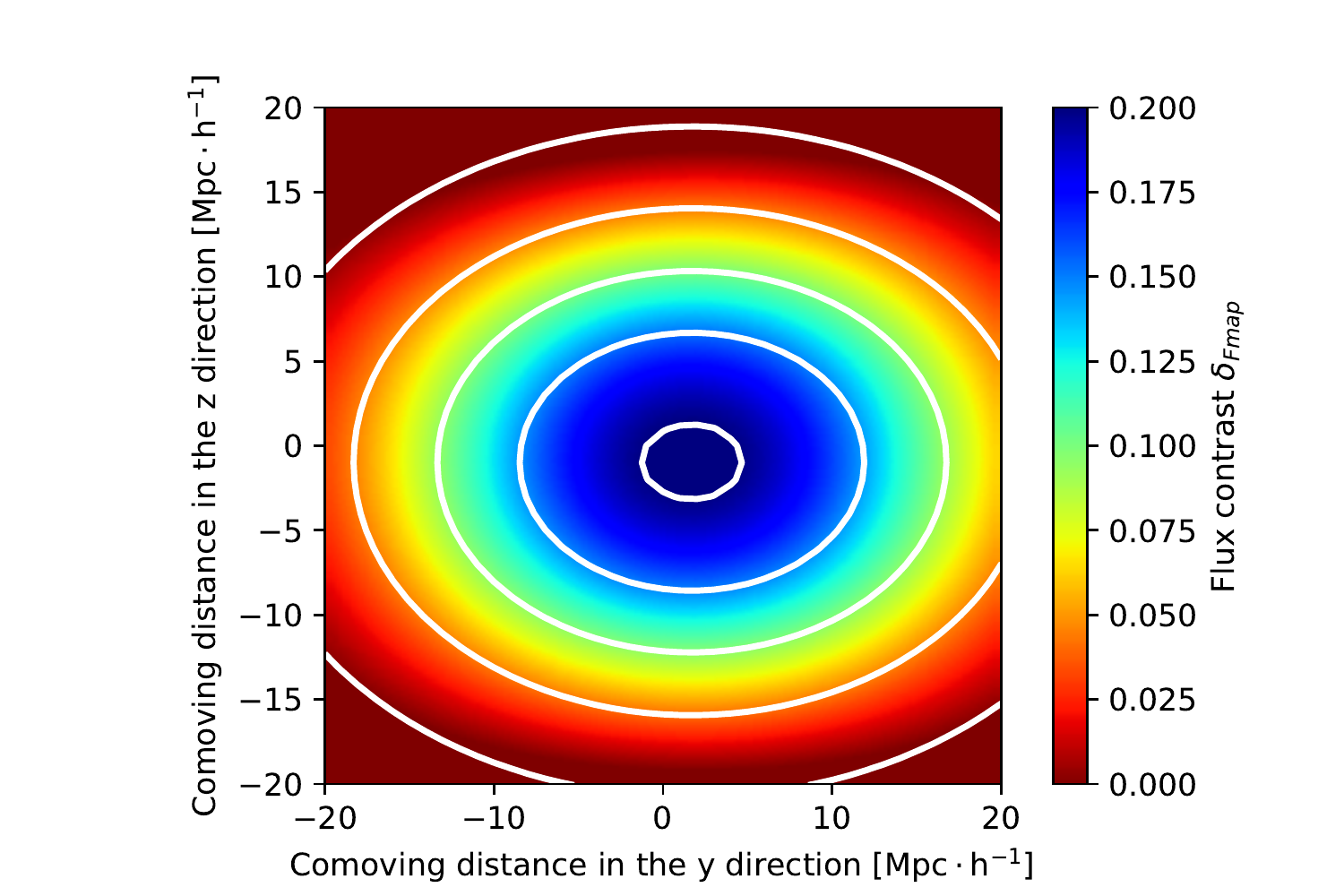}
    \caption{Stacked profile of voids identified in the mock data, with radii larger than $13\,\mpc$. The stack is over all the Stripe 82 field, with RSD effects included. White ellipses represent contours of a Gaussian fit to the profile.}
    \label{fig:stack_voids_x_RSD}
\end{figure}

It is possible to estimate the average shape of the reconstructed \lya~forest flux within voids. As smaller voids are likely to be more contaminated by noise fluctuations, we focus on voids with reconstructed radius larger than $L_{\bot} = 13\,\mpc$. From the mock tomographic map, we compute an average 3-dimensional void profile by stacking the map around all identified void centers, using the appropriate comoving coordinate system. The result is illustrated in Fig.~\ref{fig:stack_voids_x_RSD}. Contours are smooth due to the large number of voids detected over a near-Gpc$^3$ volume. In the $x-y$ plane (perpendicular to the lines-of-sight) the void profile has a circular shape. However, as shown in Fig.~\ref{fig:stack_voids_x_RSD}, it is clearly anisotropic in the $y-z$ plane: voids are more extended perpendicularly to the lines-of-sight. To quantify this effect, we fit the stacked void profile by a 2-dimensional Gaussian function. Axis ratios in the $x-y$, $y-z$ and $x-z$ planes are defined by $d_{ij} = \sigma_{i}/\sigma_{j}$ where $\sigma_{k}$ is the standard deviation of the 2D Gaussian along the axis $k$. Statistical error bars on these $d_{ij}$ are computed with the Jackknife resampling method. The measured axis ratios are $d_{yz} = 1.37 \pm 0.01$ and $d_{xz} = 1.36 \pm 0.01$. As a reference, the axis ratio in the $x-y$ plane is $d_{xy} = 0.98 \pm 0.005$. 

There are three plausible causes to this $\sim 36$~\%-level anisotropy:

\begin{itemize}
    \item An error in the choice of coordinate system, or equivalently in the underlying cosmological model, can generate an anisotropy in the average void profile - the Alcock-Paczynski effect \cite{Alcock1979}. Unless drastically departing from the coordinate system implied by existing contraints on cosmological parameters (\cite{Planck2018}), such an effect cannot explain the observed axis ratio. We also checked that our use of conventional $(x,y,z)$ coordinate system, described in Sect.~\ref{sec:Algo}, instead of true comoving coordinates, does not modify the observed axis ratio.
    \item Redshift-space distortions (RSD) are present in the \lya~forest flux, and taken into account in mocks by the term $c(z)\,\eta_{\parallel}$ in Eqn.~\ref{eq:flux_mocks_2}. To assess their impact on the average mock void profile, we computed a dedicated mock where the term $c(z)\,\eta_{\parallel}$ is suppressed. Its axis ratio is $d_{yz} = 1.33$, smaller than when RSD are included ($d_{yz}=1.37 \pm 0.01$). This result demonstrates that RSD do impact the observed void anisotropy, but are not a dominant factor.
    \item While the Wiener filter algorithm we adopted is isotropic in itself ($L_{\parallel}=L_{\bot}$), the input distribution of line-of-sight pixels is strongly anisotropic, with straight, parallel lines-of-sight. The tomographic map inherits this anisotropy. For example, by artificially subsampling the input line-of-sight mock data, the statistical isotropy of the corresponding pixels is increased and the mean void axis ratio is decreased. This behavior indicates that a large fraction of the mean void axis ratio is due to this effect.
\end{itemize}



\section{Results on eBOSS data}
\label{sec:Results}

\subsection{Tomographic map and properties of voids}

\begin{figure}[!t]
    \centering
    \includegraphics[trim=2.5cm 1cm 2cm 2cm, clip=true,width = 165mm]{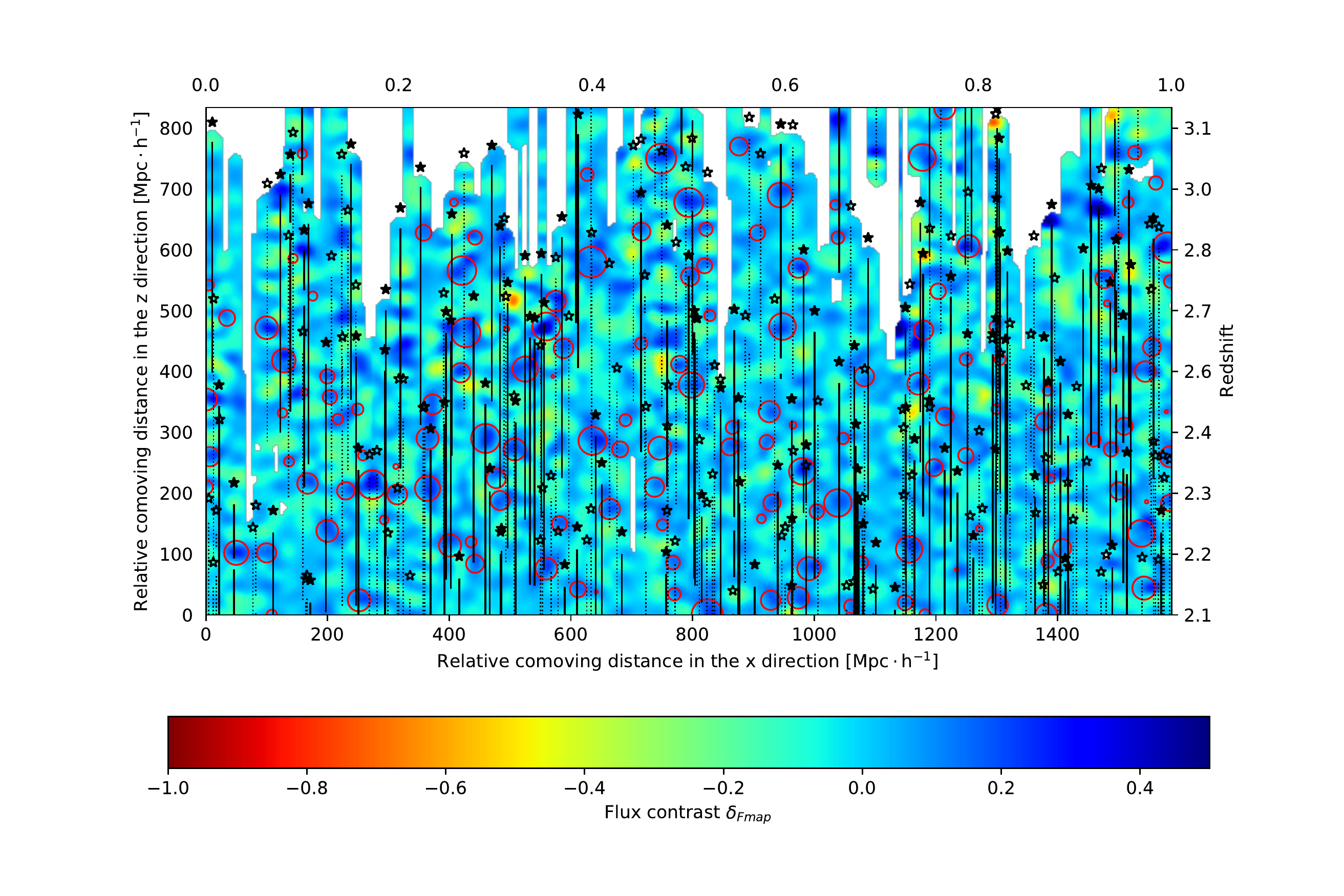}
    \caption{Slice of the tomographic map computed from the observed eBOSS \lya~forest fluxes in the Stripe 82. The slice is at constant declination $\delta_{J2000} = 0\degree$, and covers RA $\in [1\degree,23\degree]$, i.e., roughly a quarter of the Stripe 82 field. The lines-of-sight, quasars, and voids are represented as in Fig.~\ref{fig:map_saclay_mocks}. All pictured quasars and line-of-sights (full and dotted) are located in a $20\mpc$-thick slice. This thickness corresponds to $0.28\degree$ in the declination direction.}
    \label{fig:map_DR16}
\end{figure}

We apply the algorithm detailed in Sect.~\ref{sec:Algo} on the selected eBOSS - DR16 dataset presented in Sect.~\ref{sec:Stripe}, with the same parameters and output format as those for the mock map presented in Sect.~\ref{sec:Mocks}. In particular, pixels located more than $20\,\mpc$ from any given line-of-sight are masked.The resulting map is available on \texttt{Zenodo}\footnote{\label{zenodo_data_release}\url{https://doi.org/10.5281/zenodo.3737781}}. Fig.~\ref{fig:map_DR16} presents a constant-declination slice of this map over one quarter of the Stripe 82 field. 

\begin{figure}[!t]
    \centering
    \includegraphics[trim=0cm 0cm 0cm 0cm, clip=true,width = 165mm]{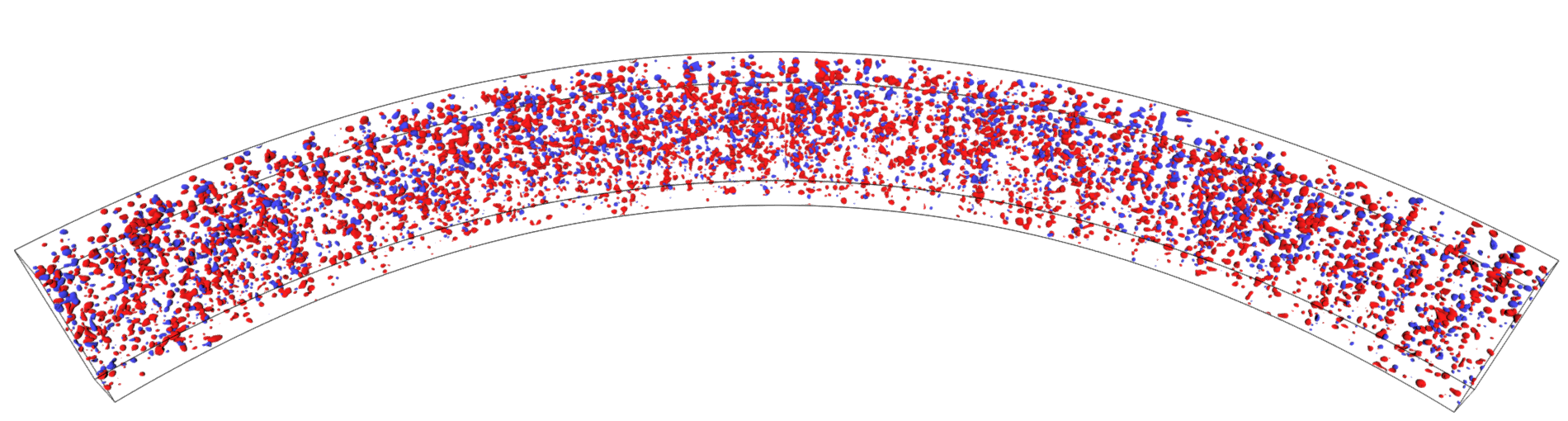}
    \caption{3D representation of the tomographic map over the entire Stripe 82, in spherical coordinates. Red (blue) iso-surfaces of flux contrast $-0.2$ ($0.2$) represent overdensities (underdensities). The corresponding 3D interactive representation is available on: \url{https://skfb.ly/6RyGL}. An interactive representation of overdensities is available on: \url{https://skfb.ly/6RyGE}.}
    \label{fig:3D_map}
\end{figure}

\begin{figure}[!t]
    \centering
    \includegraphics[trim=0cm 0cm 0cm 0cm, clip=true,width = 150mm]{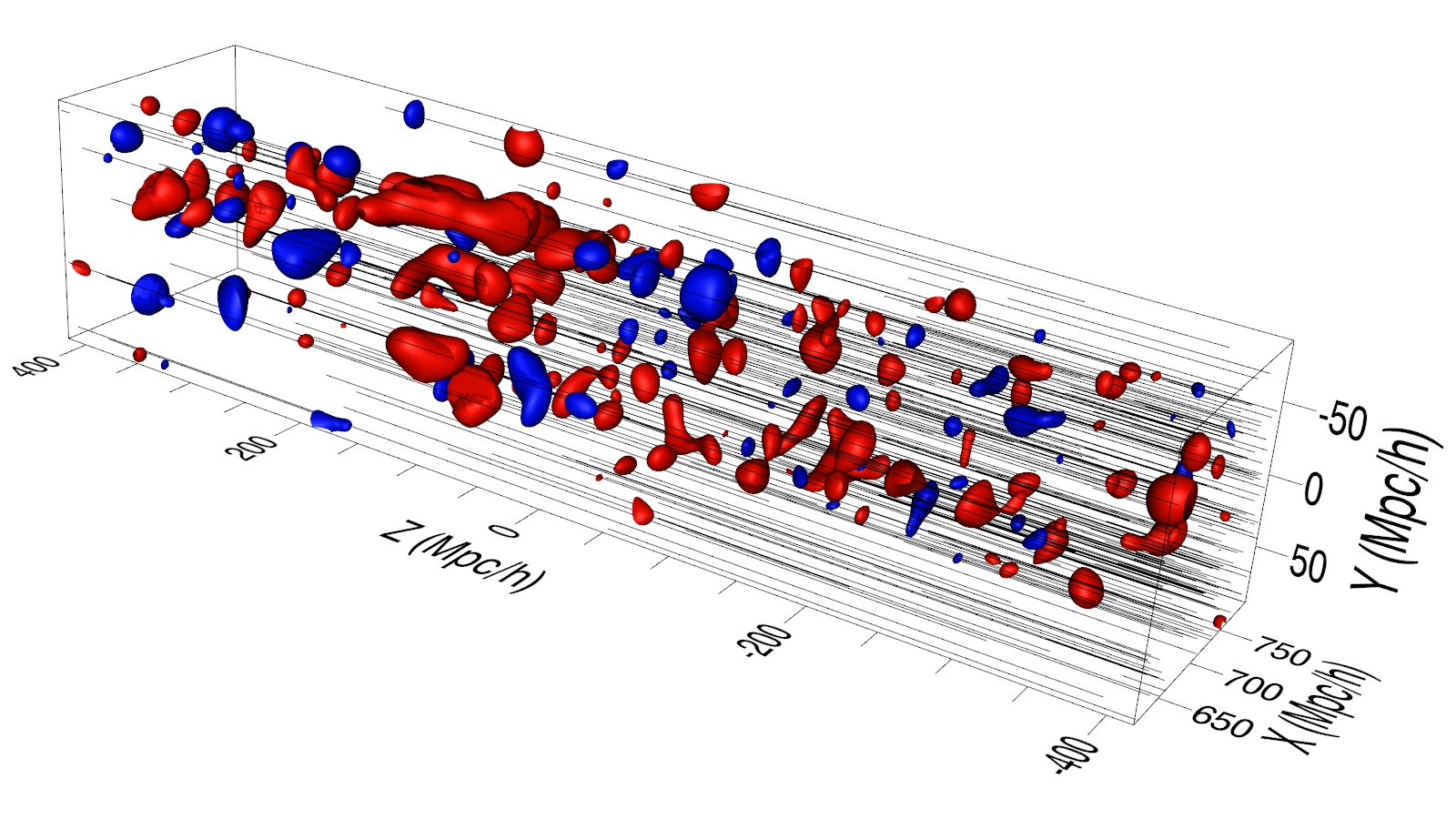}
    \includegraphics[trim=0cm 0cm 0cm 0cm, clip=true,width =120mm]{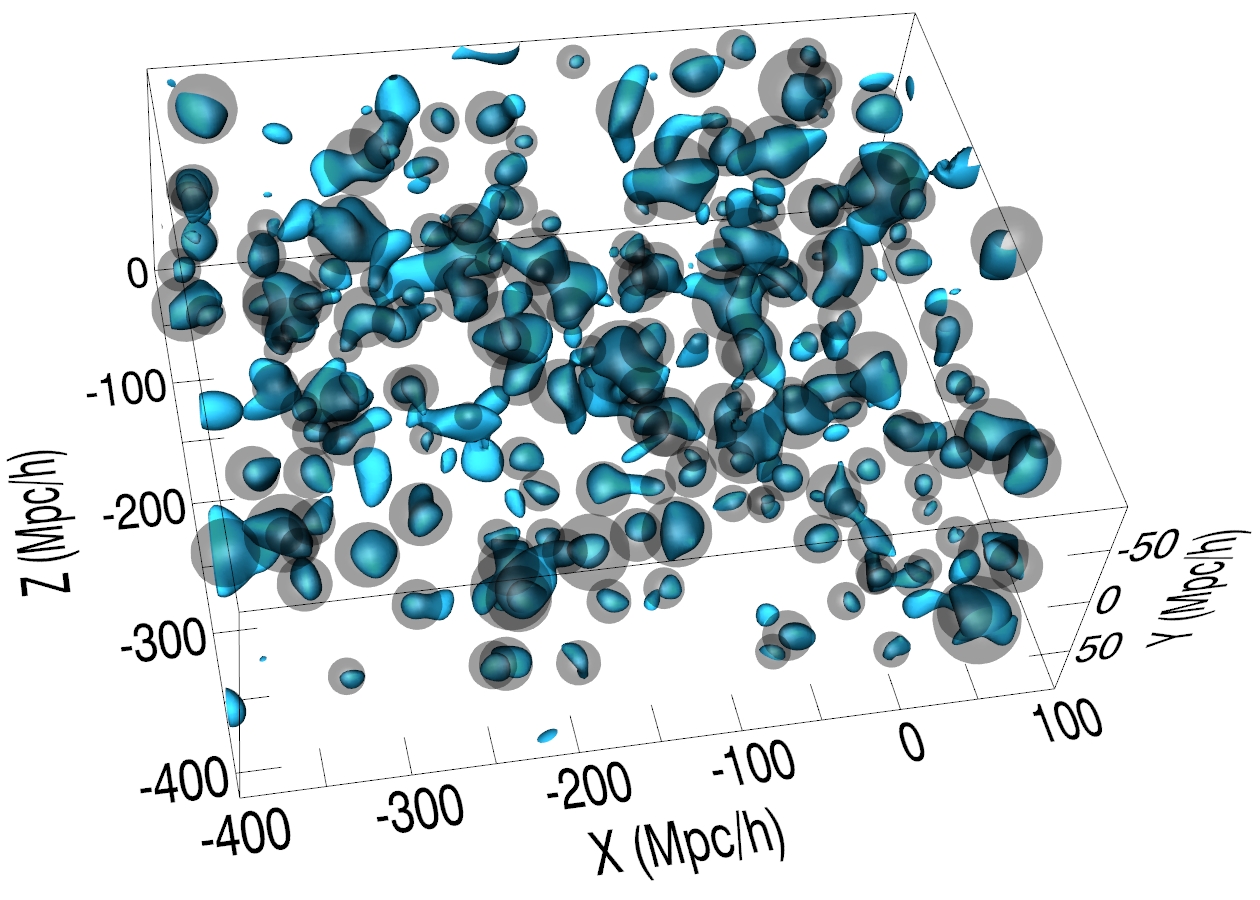}

    \caption{3D representations of portions of the Stripe 82 field. Top: $2.5\degree\times 2.5\degree$ field in right ascension and declination. Red (blue) iso-surfaces of flux contrast $-0.2$ ($0.2$) represent overdensities (underdensities). Lines-of-sight are represented in black. Bottom: $7\degree\times 2.5\degree$ field in right ascension and declination. Blue iso-surfaces of flux contrast 0.14 represent underdensities, at the same level than searched by the void finder ($\delta_{th}$). Identified voids are represented by light grey spheroids. The corresponding 3D interactive representations are available on \url{https://skfb.ly/6RyGn} (top) and \url{https://skfb.ly/6RyGP} (bottom).}
    \label{fig:3D_map_LOS_voids}
\end{figure}

In addition we provide 3D representations of the tomographic map. Some visualizations were prepared with the Saclay Data Visualization (SDvision) \cite{Pomarede2017} software deployed in the IDL environment, and also provide interactive web-based visualizations hosted on the Sketchfab\footnote{\label{sketchfab}\url{https://sketchfab.com/}} platform. The SDvision software allows for the interactive and immersive visualization of scalar and vector fields, clouds of points, as well as ancillary datasets such as the lines-of-sight and catalogues of reconstructed voids. In our representations, the \lya~forest flux contrast is visualized using isocontour 3D surfaces at negative and positive values to indicate overdense and underdense regions, respectively. Sketchfab is a web-based service that enables the sharing of 3D models. The user can rotate, pan, zoom in and out regions of interest, and benefit from GPU acceleration through the WebGL API, as well as immersive Virtual Reality capabilities in association with VR headsets.
Examples and links to these 3D interactive animations are given in Fig.~\ref{fig:3D_map}~and \ref{fig:3D_map_LOS_voids}. All the corresponding 3D interactive animations are available on \url{https://sketchfab.com/pomarede/collections/eboss-paper}.

\begin{figure}[!t]
    \centering
    \begin{subfigure}[t]{0.5\textwidth}
        \centering
        \includegraphics[trim=0cm 0cm 0cm 0cm, clip=true,width = 72mm]{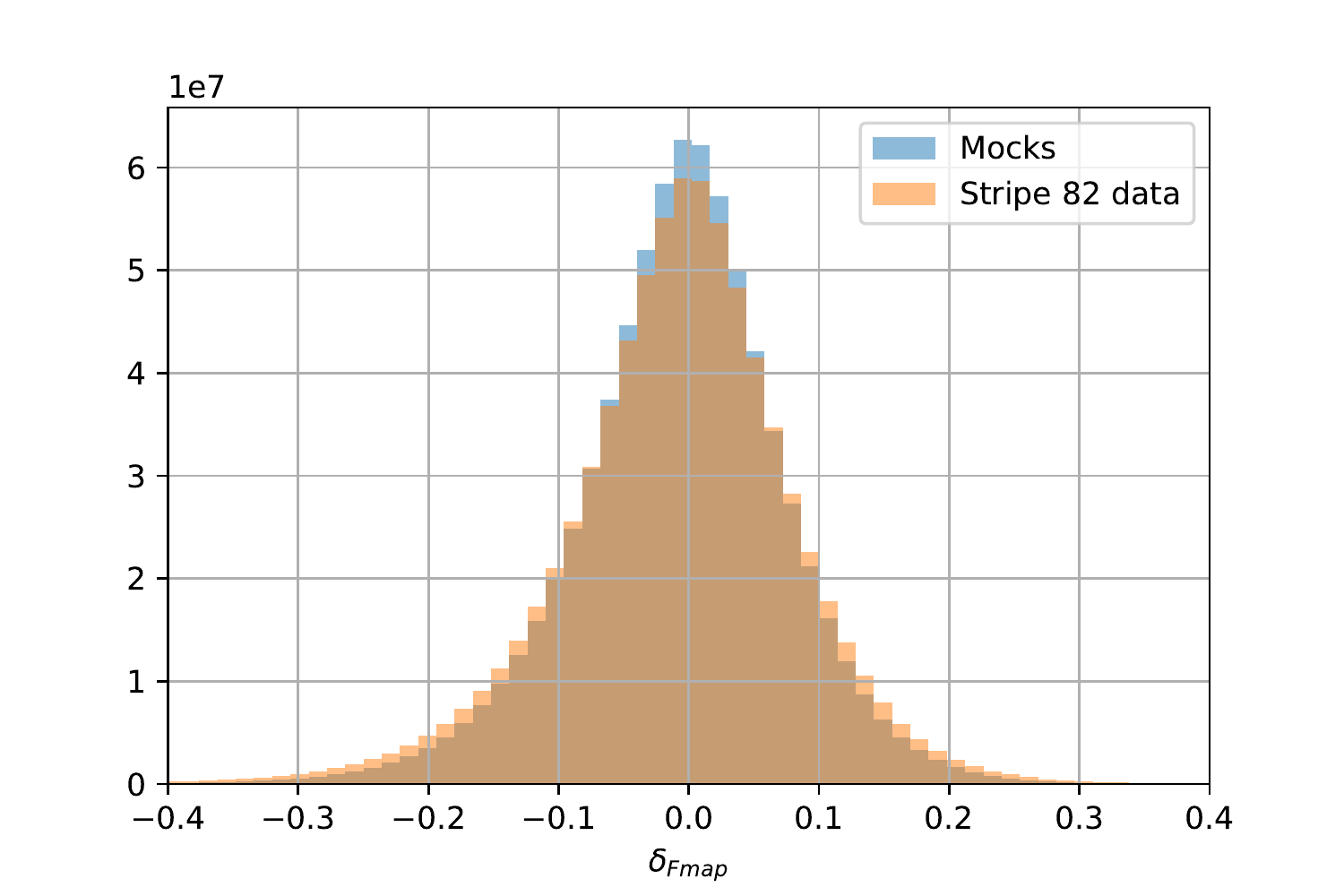}
    \end{subfigure}%
    ~ 
    \begin{subfigure}[t]{0.5\textwidth}
        \centering
        \includegraphics[trim=0cm 0cm 0cm 0cm, clip=true,width = 72mm]{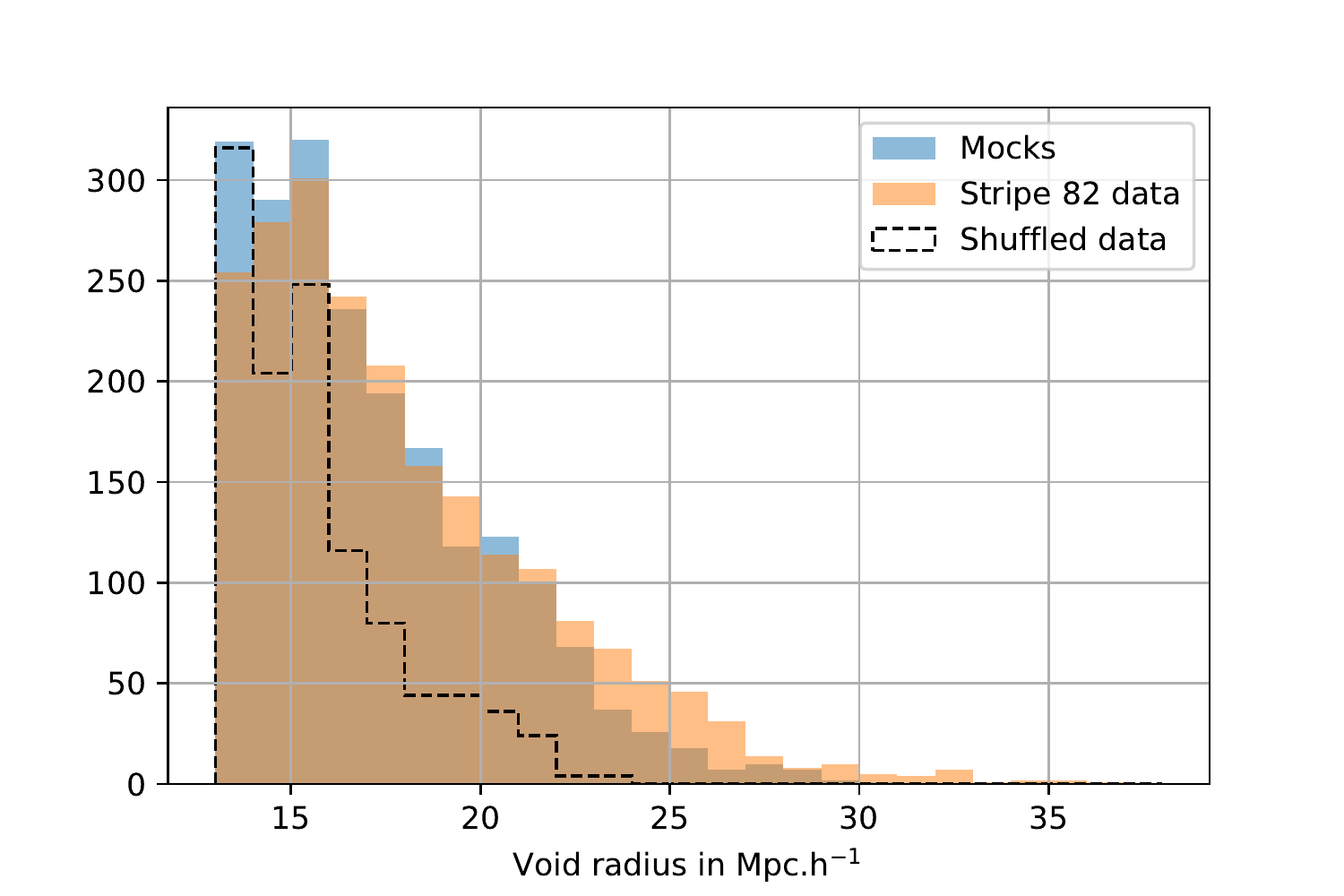}
    \end{subfigure}
    \caption{Comparison between tomographic map properties derived from the mocks and for eBOSS DR16 data. Left: Histogram of the \lya~forest flux contrast. Right: Histogram of void radii (for radii larger than $L_{\bot} = 13\,\mpc$). Dashed line represent void radii obtained after shuffling the input data, see Sect.~\ref{sec:Results}.}
\label{fig:histogram_comparison_mock_data_lin}
\end{figure}

The flux contrast distribution in the reconstructed map is presented and compared to that of the mock map in Fig.~\ref{fig:histogram_comparison_mock_data_lin} (left). Both histograms are well approximated by a Gaussian distribution; however, we observe a higher standard deviation in the data distribution than in the mocks. This result may be related to the difference in noise distribution that we pointed out in Fig.~\ref{fig:histogram_noise_comparison_mock_data}.

Large voids in the matter distribution are identified by applying the void finder algorithm on the Stripe 82 data with the same parameters as described in Sect.~\ref{subsec:void_algo}. The voids are represented by circles on the slice view of Fig.~\ref{fig:map_DR16}. A catalog providing void coordinates and dimensions is available on \texttt{Zenodo}$^{\ref{zenodo_data_release}}$. A total of 439 void candidates are detected in Stripe 82 with a radius larger than $20\,\mpc$.  The distribution of void radii is shown in Fig.~\ref{fig:histogram_comparison_mock_data_lin} (right), and compared to that of mock data. For both Stripe data and mocks, a tail of distribution is observed. The profile is well fit by an exponential law $dN/dR \sim \exp(-R/R_0)$, with $R_0 = 4.9\pm 0.3\,\mpc$ ($5.6 \pm 0.3$) for data (mocks). This feature of the distribution is expected and comparable to the measured statistical properties of low-redshift voids, e.g., \cite{Sutter2013}.


In order to investigate the physical origin of our large-void tail of distribution, we shuffled the line-of-sight pixels randomly from the initial $\delta_{F}$ data, keeping the line-of-sight positions fixed. After map-construction and void finding, the resulting void radius distribution is shown in Fig.~\ref{fig:histogram_comparison_mock_data_lin} (right, dashed line). The high-radius tail observed in data is absent after shuffling: the number of voids with a radius larger than $20\,\mpc$ is 6~\% of the one observed in non-shuffled data. For void candidates with radii $\lesssim 15\,\mpc$, the contamination from noise fluctuations is important. This result indicates that low radius voids are contaminated by false positive detections. As a comparison, in the galaxy field, random Poisson point distribution gives very similar results concerning the distribution of voids~\cite{Nadathur2015a,Nadathur2015b}.

Comparing mocks and data, a Kolmogorov-Smirnov test demonstrates that the mock and data void radius distributions are not compatible with each other, with more large voids in the data than in the mocks. This difference might arise from differences between noise distributions or the modelization of synthetic HCD systems in the mocks.


\subsection{Matter overdensities}

We now present the result of searches for matter overdensities in the tomographic map, both a blind search and searches around known objects.

\begin{figure}[!t]
    \centering
    \begin{subfigure}[t]{0.5\textwidth}
        \centering
        \includegraphics[trim=0.25cm 0cm 0cm 0cm, clip=true,width = 75mm]{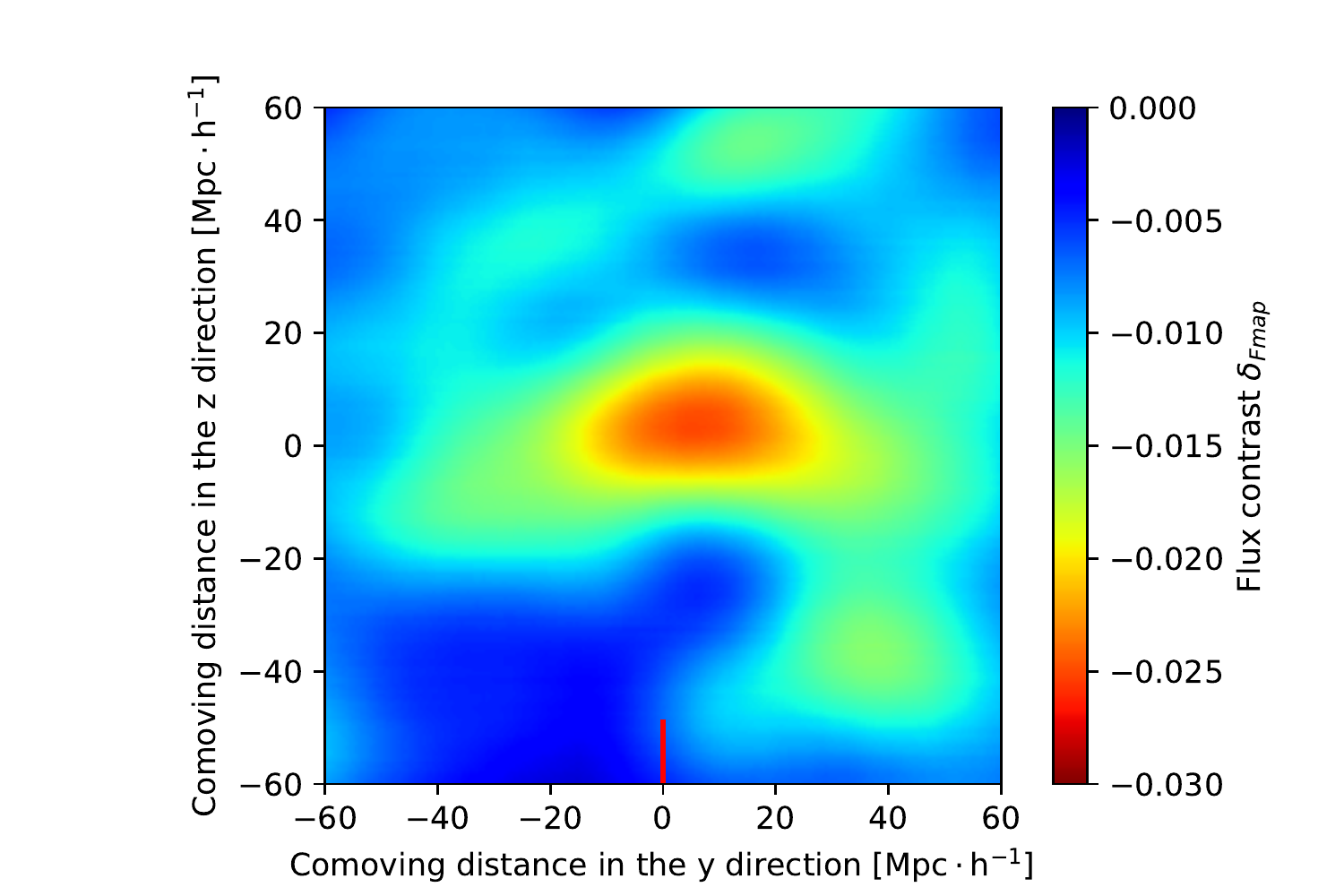}
    \end{subfigure}%
    ~ 
    \begin{subfigure}[t]{0.5\textwidth}
        \centering
        \includegraphics[trim=0.25cm 0cm 0cm 0cm, clip=true,width = 75mm]{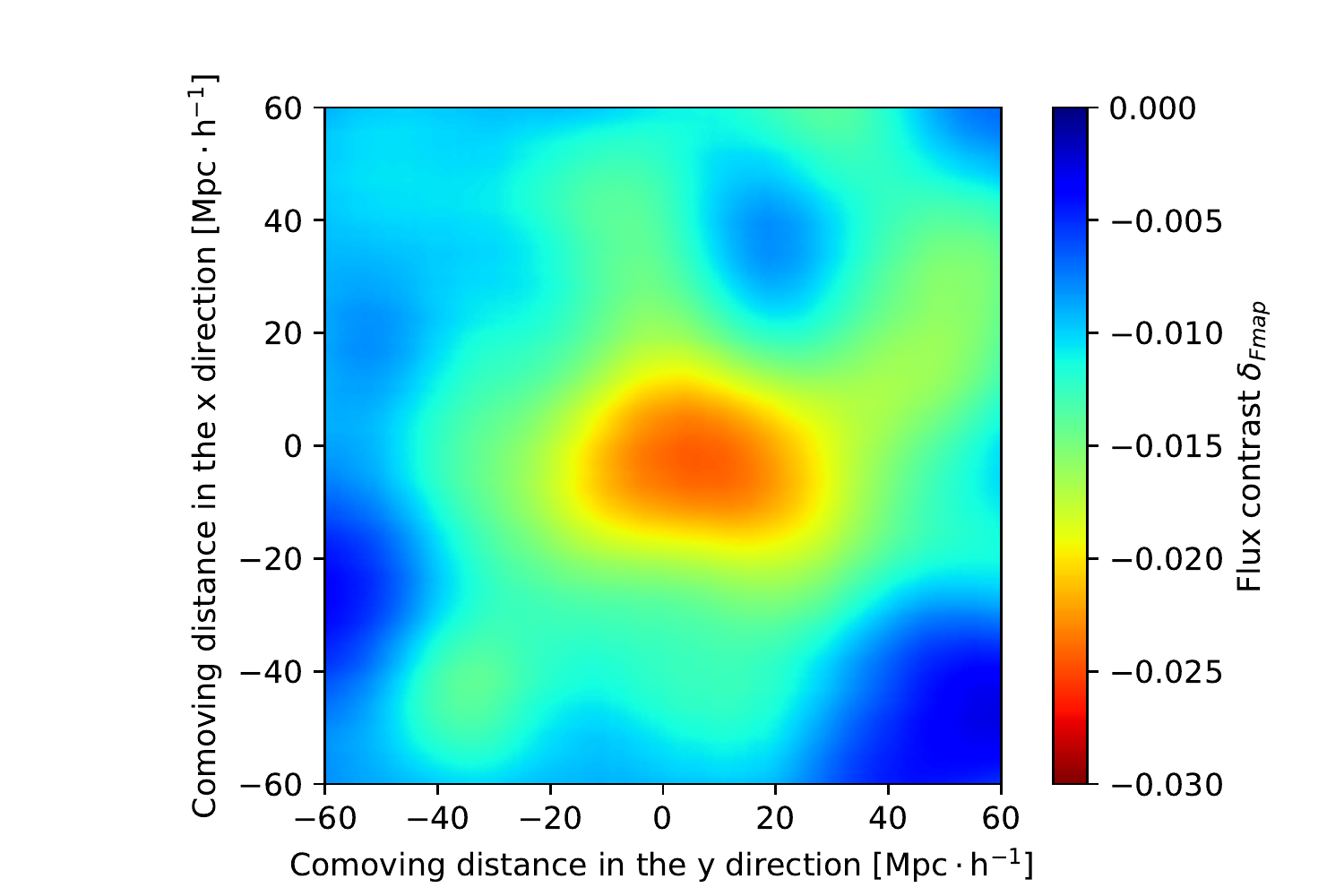}
    \end{subfigure}
    \caption{Stacked tomographic \lya~forest map centered around quasars over Stripe 82. Left: View in the $y-z$ plane. The mean line-of-sight position with respect to the quasar is represented by a red line, at the bottom of the image at $y=0$. Right: view in the $x-y$ plane.}
    \label{fig:stack_quasars}
\end{figure}

In the Stripe 82 field, the only large and homogeneous sample of known objects with $z>2.1$ and accurate (spectroscopic) redshift determination are the quasars actually used as targets for BOSS/eBOSS \lya~forest flux measurements. Fig.~\ref{fig:stack_quasars} displays our Stripe 82 tomographic \lya~forest flux stacked around quasars in the same field, with $z=2.1 - 3.2$. We observe a 3-dimensional signal, extending over roughly $20\,\mpc$. By stacking on random map positions, the central signal of this stacked map produces a $5.2\sigma$ excess with respect to a random distribution. This result is a recast view of the small-separation part of the cross-correlation between quasar positions and \lya~forest flux described in~\cite{Bourboux2017,Blomqvist2019}. The statistical significance of the signal is, however, smaller, since this study is limited to the Stripe 82 field.

Since quasars are by definition located at the end of lines-of-sight, there is a concern that the observed cross-correlation signal is biased by the Wiener filtering algorithm itself. However, the tomographic \lya~forest flux is peaked right at the quasar position, within statistical uncertainties, whereas the lines-of-sight used for the reconstruction are located on average $49\,\mpc$ away, i.e., 3 times the width $L_{\bot}$ of the filtering kernel. Such a bias, if any, is therefore a minor effect.

A hint for anisotropy in the stacked signal around quasars is visible in the $y-z$ slice in Fig.~\ref{fig:stack_quasars}. The asymmetry of the signal along the lines-of-sight is smaller than its sky-coordinate extension. Using the same method as in Sect.~\ref{subsec:underdensities_props}, we find that the axis ratios of the stacked signal are $d_{xy} =0.68 \pm 0.15$, $d_{xz} = 2.31 \pm 0.23$, and $d_{yz} = 2.52 \pm 0.63$, with statistical error bars estimated by Jackknife resampling.

We computed the same stack around quasars in the mock map; the corresponding axis ratios are $d_{xy} = 0.77 \pm 0.16$, $d_{xz} = 1.79 \pm 0.06$, and $d_{yz} = 2.18 \pm 0.10$.
The anisotropy observed in the data is therefore statistically sound, and compatible with predictions from simulations. The origin of this anisotropy may be, as in the case of voids presented in Sect.~\ref{subsec:underdensities_props}, a combination of numerical effects in the map-making procedure and RSD. The proximity effect should not be responsible for this anisotropy, since it was not taken into account in our mock sample.

We also examined the correlation of the tomographic map with DLAs. We applied the same stacking procedure as was done for quasars, centering on the positions of identified DLAs in the Stripe 82, with $N_{\rm HI} > 10^{20.3} \, \mathrm{cm^{-2}}$. There is an $2.6\,\sigma$ deficit of tomographic \lya~forest flux at the DLA positions. The signal is clearly fainter than for quasars as there is an order of magnitude less detected DLAs.

We now turn to a blind search for overdensities. This search aims at identifying candidate high-redshift protoclusters or more generally large overdensities over the wide Stripe 82 field.
As detailed in Sect.~\ref{subsec:protocluster_algo}, we apply the watershed algorithm on the Stripe 82. The adopted threshold is $\delta_{ws} = -3.5\,\Tilde{\sigma}_{\delta_{\mathrm{Fmap}}} = -0.273$, where the standard deviation of the flux contrast of the map $\Tilde{\sigma}_{\delta_{\mathrm{Fmap}}}$ is computed by using a Gaussian fit of the $\delta_{\mathrm{Fmap}}$ distribution.

These criteria yield eight selected overdensities in the tomographic map; table~\ref{tab:protocluster} lists their properties. Their average estimated radius is $12\,\mpc$; as expected, they correspond to particularly large, and massive, protocluster candidates. The number of candidate overdensities is compatible with that found from the mock map in Sect.~\ref{subsec:protocluster_algo}. Assessing in more detail the completeness and robustness of this selection would require the use of N-body simulations adapted to the identification of protocluster halos, which is beyond the scope of this article.

\begin{table}[tbp]
\begin{tabular}{|l|cccccccc|}
\hline
 & 1 & 2 & 3 & 4 & 5 & 6 & 7 & 8  \\
\hline
RA$_{J2000}$ [deg] & -37.621 & -27.442 & 3.1647 & 33.792 & 38.748 & 41.031 & 41.481 & 41.421  \\
DEC$_{J2000}$ [deg]& 0.583 & -0.166 & 0.472 & 0.750 & -1.166 & -0.0276 & -0.528 & -0.0276  \\
Redshift & 2.734 & 2.818 & 2.525 & 2.435 & 2.377 & 3.170 & 2.111 & 2.764  \\
Radius [$\mpc$]& 15.13 & 10.54 & 13.88 & 10.68 & 9.081 & 7.623 & 16.20 & 12.77  \\
Radius [arc min.] & 12.56 & 8.75 & 11.52 & 8.869 & 7.539 & 6.329 & 13.45 & 10.60  \\
Minimal $\delta_{\mathrm{Fmap}}$ & -0.76 & -0.44 & -0.69 & -0.63 & -0.38 & -0.40 & -0.74 & -0.54  \\
\hline
\end{tabular}
\caption{Properties of selected overdensities (protocluster candidates) over all the volume of our Stripe 82 tomographic map.}
\label{tab:protocluster}
\end{table}

\section{Conclusion}
\label{sec:Conclusion}

From the BOSS and eBOSS quasar spectra measured within the Stripe 82 field, we constructed a near-Gpc$^3$ volume 3D tomographic map of the \lya~forest flux contrast. Its $13\,\mpc$ resolution is such that it directly maps the large-scale distribution of matter, both baryonic and dark. This map is a unique representation of matter fluctuations at high redshifts $z>2$ over such a volume.

From synthetic data, we evaluate a $-34\%$ correlation coefficient of the tomographic map with respect to the underlying matter fluctuations, smoothed over $8\,\mpc$. This value is comparable to other estimations based on N-body simulation~\cite{Ozbek2016}. 

We presented a search for large voids over the volume of the map. There is a reasonable match between the statistical properties of voids in mocks and Stripe 82 data; in particular, we unveil a tail of distribution of voids in comoving radius, reaching up to $\sim 30\, \mpc$. The void geometry in the tomographic map is not isotropic. While mocks demonstrate that a small fraction of their axis ratio originates from RSD, most of it arises from the line-of-sight geometry of \lya~forest data.

Stacking the tomographic map around quasars yields a clear cross-correlation signal, with amplitude and shape in agreement with simulations including the RSD effect. The map is also well adapted to protocluster searches with radius larger than $7\,\mpc$; we provide a list of candidates identified by a watershed algorithm.

Unlike the CLAMATO map \cite{Lee2014,Lee2018}, the SDSS line-of-sight density is not sufficient to investigate the filamentary structure of the cosmic web. However, with the explored volume, this SDSS tomographic map allows the identification of potential large cosmic voids and overdensities with unprecedented statistical power. Compared to weak-lensing tomography, e.g., \cite{Oguri2017}, \lya~forest tomographic maps can better estimate the matter distribution in the longitudinal direction, which is limited by the width of the lensing kernel in the case of weak lensing.

There are several avenues for improvement, both in terms of methodology and data. First, more advanced algorithms can be used, in particular following the forward modelling approach \cite{Porqueres2019,Horowitz2019}. Other technical improvements are also possible, such as including the Lyman-$\beta$ region of the spectra.

Cross-correlation studies with other tracers are currently limited given the high redshift range of our analysis. Ongoing and future surveys will address this issue, with applications, for example, in the search for protoclusters. The tomographic map in itself could be used for several cosmological analysis different from the typical correlation function-based approach, e.g., advanced shape analysis of over- and under-densities could lead to other cosmological constraints.

Finally, while this study focuses on the Stripe 82 field as observed with SDSS data, future work will be dedicated to studies of other fields, in particular making use of the upcoming DESI~\cite{Desi2016} and WEAVE-QSO~\cite{Pieri2016} surveys. For these two surveys, mean line-of-sight separation down to $7\,\mpc$ can be achieved in very dense regions, and will lead to more extended and more refined tomographic maps.


\acknowledgments

Funding for the Sloan Digital Sky Survey IV has been provided by the Alfred P. Sloan Foundation, the U.S. Department of Energy Office of Science, and the Participating Institutions. SDSS-IV acknowledges support and resources from the Center for High-Performance Computing at the University of Utah. The SDSS web site is www.sdss.org.

SDSS-IV is managed by the Astrophysical Research Consortium for the Participating Institutions of the SDSS Collaboration including the Brazilian Participation Group, the Carnegie Institution for Science, Carnegie Mellon University, the Chilean Participation Group, the French Participation Group, Harvard-Smithsonian Center for Astrophysics, Instituto de Astrof\'isica de Canarias, The Johns Hopkins University, Kavli Institute for the Physics and Mathematics of the Universe (IPMU) / University of Tokyo, the Korean Participation Group, Lawrence Berkeley National Laboratory, Leibniz Institut f\"ur Astrophysik Potsdam (AIP), Max-Planck-Institut f\"ur Astronomie (MPIA Heidelberg), Max-Planck-Institut f\"ur Astrophysik (MPA Garching), Max-Planck-Institut f\"ur Extraterrestrische Physik (MPE), National Astronomical Observatories of China, New Mexico State University, New York University, University of Notre Dame, Observat\'ario Nacional / MCTI, The Ohio State University, Pennsylvania State University, Shanghai Astronomical Observatory, United Kingdom Participation Group, Universidad Nacional Aut\'onoma de M\'exico, University of Arizona, University of Colorado Boulder, University of Oxford, University of Portsmouth, University of Utah, University of Virginia, University of Washington, University of Wisconsin, Vanderbilt University, and Yale University.

The authors acknowledge support from grant ANR-16-CE31-0021. In addition, this research relied on resources provided to the eBOSS Collaboration by the National Energy Research Scientific Computing Center (NERSC).  NERSC is a U.S. Department of Energy Office of Science
User Facility operated under Contract No. DE-AC02-05CH11231.

\bibliographystyle{JHEP.bst}
\bibliography{bibli}

\end{document}